\documentclass[apj,twocolumn]{openjournal}
\usepackage[utf8]{inputenc}
\usepackage{amsmath}
\usepackage{hyperref}
\usepackage{yhmath}
\newcommand{\mi}{\mathrm{i}}
\newcommand{\md}{\mathrm{d}}
\newcommand{\cm}{\checkmark}
\newcommand{\crs}{\boldsymbol{\times}}
\newcommand{\bb}[1]{\mathbf{#1}}
\usepackage{color}
\usepackage[T1]{fontenc}  
\usepackage{aas_macros}

\usepackage{comment}
\usepackage{cancel}
\usepackage{natbib}
\setcitestyle{numbers}

\begin{document}
\rightline{\scriptsize RBI-ThPhys-2025-12}

\title{The Bispectrum of Intrinsic Alignments: Theory Modelling and Forecasts for Stage IV Galaxy Surveys}
\author{Thomas Bakx$^{\dagger,1}$}
\author{Alexander Eggemeier$^2$}
\author{Toshiki Kurita$^{3,4}$}
\author{Nora Elisa Chisari$^{1,5}$}
\author{Zvonimir Vlah$^{6,7,8}$}
\email{$\dagger$ t.j.m.bakx@uu.nl}
\affiliation{$^1$Institute for Theoretical Physics, Utrecht University, Princetonplein 5, 3584 CC, Utrecht, The Netherlands.}
\affiliation{$^2$Universität Bonn, Argelander-Institut für Astronomie, Auf dem Hügel 71, 53121 Bonn, Germany}
\affiliation{$^3$Max-Planck-Institut für Astrophysik, Karl-Schwarzschild-Str. 1, 85748 Garching, Germany}
\affiliation{$^4$Kavli Institute for the Physics and Mathematics of the Universe (WPI), \\
The University of Tokyo Institutes for Advanced Study (UTIAS), The University of Tokyo, Chiba 277-8583, Japan}
\affiliation{$^5$Leiden Observatory, Leiden University, PO Box 9513, 2300 RA Leiden, the Netherlands.}
\affiliation{$^6$Division of Theoretical Physics, Ruđer Bo\v{s}kovi\'c Institute, 10000 Zagreb, Croatia,}%
\affiliation{$^7$Kavli Institute for Cosmology, University of Cambridge, Cambridge CB3 0HA, UK}%
\affiliation{$^8$Department of Applied Mathematics and Theoretical Physics, University of Cambridge, Cambridge CB3 0WA, UK.}

\begin{abstract}
We present a complete treatment of the bispectrum of intrinsic alignments, both in three spatial dimensions and in projection in the flat-sky approximation. Since intrinsic alignment is a spin-2 observable, the bispectrum of intrinsic alignments contains a parity-even and a parity-odd part, the latter being nonzero even in the absence of parity violation. Moreover, all possible combinations of scalar, E- and B-mode bispectra are nonzero in the absence of parity violation. In analogy to the galaxy bispectrum in redshift space, we construct a complete set of multipoles for anisotropic bispectra of projected spin-2 fields. 
We then construct separable bispectrum estimators, both for parity-even and parity-odd bispectra, which can be computed by means of Fast Fourier Transforms (FFTs). We compare several different choices of angular weighting in terms of signal-to-noise ratios (SNR) for a Stage IV setup using luminous red galaxies (LRGs) from the Dark Energy Spectroscopic Instrument (DESI) with galaxy shapes measured by the Legacy Survey of Space and Time (LSST). Assuming an overlapping area of $\sim 4,000$ square degrees (yielding $\sim 1.3$ million LRGs) and including scales up to $k_\text{max} = 0.14\,h$/Mpc, we find that the position-position-E-mode bispectrum $B_{DDE}$ (which is parity-even) can be strongly detected at SNR $\sim 30$, while detecting parity-odd bispectra (such as $B_{DDB}$, SNR $\sim 5$) or bispectra with more than one shape field (such as $B_{DEE}$, SNR $\sim 5$) may also be possible.
\end{abstract}

\maketitle

\section{Introduction}

Over the past years, precision cosmology has painted an increasingly accurate depiction of the Universe through large-scale structure surveys \cite{alam_boss,desisb}, weak lensing surveys \cite{des_kp, secco,asgari21,hsc1,hsc2}, data from nearby supernovae at low redshifts \cite{des_sn} as well as early-Universe measurements from the {\it Planck} satellite \cite{planck2018}. Simultaneously, several tensions have appeared between these probes within the concordance $\Lambda$CDM model (see \cite{Efstathiou25} for a review). As such, it is paramount that next-generation surveys with yet smaller error budgets reconsider all possible sources of systematics that may be relevant at the level of accuracy that is aimed for. On the other hand, the projected increase in signal-to-noise for future equivalents of these surveys will also enable new cross-validation and calibration methods to ensure robustness of their final results.

One prominent example of such systematics is found in the intrinsic alignments (IAs) of galaxies, which arise due to their gravitational interaction with the local tidal field \cite{troxel_ishak, joachimi_review}. As such, intrinsic alignments are correlated with large-scale structure and thus dependent on cosmology. Intrinsic alignments have long been recognized as a possible source of systematics in weak lensing surveys \cite{catelan_la}. However, it was recognized later on that intrinsic alignments could, in parallel, provide information about the early Universe \cite{kurita_png,Chisari16}, late-time growth \cite{taruya_gr} and many other cosmological and astrophysical phenomena \cite{akitsu_magn,akitsu_png,schmidt_shear,taruya_ia,okumura_gr,harvey, shiraishi}. As another interesting example, understanding the intrinsic alignment signal present in spectroscopic clustering samples is crucial when examining selection effects as a contaminant to redshift-space distortion measurements of the growth rate \cite{lamman_rsd,hirata_gr,krause_bisp}. In addition they also provide important information on galaxy formation and evolution \cite{bhowmick}

Recently, there has been an increased interest in using  beyond two-point statistics for cosmological analyses \citep[e.g.,][]{krause_beyond,halder,philcox_boss,damico_bossbisp,ivanov_precision,burger} in order to extract the maximum amount of information from large-scale structure datasets. Efficient computational methods have now brought the application of three- and four-point statistics within reach for current and next-generation surveys \cite{philcox_window, scoccimarro_fast, slepian_3pcf, shear_3pt3}. Additionally, higher-order statistics are more sensitive to parity-violating phenomena \cite{philcox_4pt,slepian_4pt}, which have gained attention when a hint of parity violation appeared from the analysis of the galaxy four-point correlation function of the SDSS DR12 sample (though see \cite{philcox_reanalysis,krolewski_boss}). 

Motivated by these developments, this work considers the bispectrum of intrinsic alignments as a cosmological observable. We are not the first to consider three-point statistics of intrinsic alignments. Some early modelling appeared in \cite{krause_bisp, troxel_sc1, troxel_sc2,semboloni}. Additionally, \cite{pyne_illustris} considered the IA bispectrum for certain specific (e.g. equilateral or isosceles) configurations in the IllustrisTNG hydrodynamical simulation, employing a phenomenological modification of the non-linear alignment (NLA) model \cite{bridle_king, pyne_forecast} combined with an effective second order kernel \cite{gilmarin_bisp_fit}. This model is not necessarily restricted to perturbative scales, since information from the 1-halo regime is included in the effective kernels. Moreover, a recent application of an alternative phenomenological model to SDSS-LOWZ data \cite{linke_lowz} yielded a good fit to the measured intrinsic alignment signal as quantified by three-point aperture mass statistics, which can be expressed as line-of-sight integrals of the full three-dimensional bispectrum. As a consequence of employing the Limber approximation, this model only depends on the bispectrum for triangles in the sky plane. Additionally, as pointed out by \cite{kilbinger_ap}, the use of equal aperture radii implies that this statistic is mostly sensitive to equilateral configurations. However, as already noted in \cite{singh}, more information can be extracted from the IA signal if the radial structure of IA correlations is used. 

Our line of work thus aims to assess the maximum possible information content in \textit{fully anisotropic} intrinsic alignment bispectra. On the modelling side, we restrict ourselves to perturbative scales ($k<0.25\,h$/Mpc), but we examine \textit{all possible triangle configurations}. Specifically, our model is grounded in the Effective Field Theory of Large Scale Structure (EFT of LSS) \cite{baumann_eft,senatore_eft} and described in more detail below. This is a robust framework for describing the clustering of tracers of the dark matter density field on large scales, where baryonic effects and nonlinearities can safely be marginalized over at the expense of introducing a modest number of additional free parameters \cite{senatore_bias,angulo_bias}. We will consider the full three-dimensional bispectrum and decompose the angular structure into multipoles, thus keeping all information encoded in the anisotropic bispectrum \cite{gualdi_bisp}. Some earlier work in this direction also appeared in \cite{schmitz_bisp,vlah_eft1,vlah_eft2}. 

Our theoretical formalism and forecast thus assume that three-dimensional positions of galaxies are available through spectroscopically measured redshifts. We thereby defer all applications in the context of photometric surveys (thus including weak gravitational lensing) to future work. Our paper is structured as follows: in Section \ref{sec:theory}, we establish the general theoretical framework employed here for the bispectrum theory model. Then in Section \ref{sec:proj}, we state our definitions for the multipole observables we consider, as well as the construction of separable estimators. We conclude this Section by visualizing the IA bispectrum signal across all triangle configurations. In Section \ref{sec:forecast} we describe our forecasting setup for the signal-to-noise ratio (SNR) of the bispectrum multipoles and discuss the theoretical Gaussian covariance model. Section \ref{sec:results} shows a detailed account of our results. We discuss them at length in Section \ref{sec:disc}. Some calculations and results thereof are relegated to appendices.  

Throughout this work, we use the convention that the Fourier transform and inverse Fourier transform of any function $f(\bb{x})$ are 
\begin{equation}
\begin{aligned}
    f(\bb{k}) &= \int \md^3 \bb{x}\, f(\bb{x})e^{\mi \bb{k}\cdot \bb{x}}; \nonumber \\ 
    f(\bb{x}) &= \int \frac{\md^3 \bb{k}}{(2\pi)^3} \,e^{-\mi \bb{k}\cdot \bb{x}}f(\bb{k}),
    \end{aligned}
\end{equation}
respectively. Throughout the entire paper, we use $c=1$. The fiducial cosmology we use for the theory and covariance computation is Euclidean $\Lambda$CDM with $\omega_b=0.02225, \omega_c=0.1198,\Omega_\Lambda = 0.6844, n_s=0.9645,\ln(10^{10}A_s)=3.094$, identical to that of \cite{kurita_iapower} and in accordance with {\it Planck} CMB data \cite{planck2018}. We use the Boltzmann solver \texttt{CAMB} \cite{camb}.

\section{Effective Field Theory for Galaxy Shapes}\label{sec:theory}
In the past decade, it has become clear that galaxy clustering properties in the weakly non-linear regime can be described accurately by means of perturbative methods, most notably the effective field theory of large-scale structure (EFT of LSS) \cite{baumann_eft,senatore_eft}. This framework extends standard perturbation theory (SPT) methods to correctly take into account the backreaction of small scales onto the perturbative regime, as well as nonlocal processes that influence tracer formation and evolution (see \cite{schmidt_bias} for a review). Originally, this was applied to scalar tracers of the dark matter distribution, but it is also possible to generalize this to tensorial tracers such as halo shapes or galaxy shapes. This is referred to as the effective field theory of intrinsic alignments (EFT of IA) \cite{vlah_eft1,vlah_eft2}. 
\subsection{EFT Expansion for Shape Fields} Following \cite{vlah_eft1,bakx_eft} the EFT of IA describes the $N$-point statistics of a symmetric 2-tensor field $S_{ij}(\mathbf{x})$ called the shape perturbation. It is formally constructed by modelling the 3D shape of a galaxy (or dark matter halo, or any tracer of the dark matter field) located at position $\bb{x}_\alpha$ through its second moment tensor $I_{ij}(\bb{x}_\alpha)$: 
\begin{equation}
I_{ij}(\bb{x}_\alpha)\propto \sum_{p\in \alpha} w_p \Delta{\bf x}_{p,i}\Delta{\bf x}_{p,j}. \label{eq:Iij}
\end{equation}
Here $\Delta\bb{x}_{p,i}$ is difference of the $i$-th coordinate of the $p$-th particle and the center $\bb{x}_\alpha$. Different weightings $w_p$ have been employed in the literature. One can now formally construct the `moment tensor field' $I_{ij}(\bb{x})$:
\begin{equation}
    I_{ij}(\mathbf{x}):=\sum_\alpha I_{ij}({\bf x}_\alpha)\delta^D({\bf x}-{\bf x}_\alpha).
\end{equation}
The shape perturbation is subsequently defined as
\begin{equation}\label{shapedef}
    S_{ij}({\bf x}) := \frac{I_{ij}({\bf x})-\langle I_{ij} \rangle}{\langle\text{Tr}(I_{ij})\rangle} = \frac{1}{3}\delta_{ij}\delta_s(\bb{x}) + g_{ij}(\bb{x}).
\end{equation}
Here $\delta_s(\bb{x})$ is the intrinsic size perturbation (i.e. it is not lensing-induced) and $g_{ij}(\bb{x})$ is the intrinsic shape perturbation, which is traceless and has five degrees of freedom. Note that $S_{ij}$ is a dimensionless quantity and by construction $\langle S_{ij} \rangle = 0$. By virtue of statistical isotropy, we must have that $\langle I_{ij}\rangle \propto \delta_{ij}^K$.
Any symmetric tensor field $S_{ij}(\mathbf{x})$ can be decomposed into a trace and trace-free part: 
\begin{equation}
    S_{ij}(\mathbf{x}) = \frac{1}{3}\delta_{ij}\text{Tr}(S)(\mathbf{x}) + \text{TF}(S)_{ij}(\mathbf{x}).
\end{equation}
This decomposition is linear, and thus also applies in Fourier space:
\begin{equation}
    S_{ij}(\mathbf{k}) = \frac{1}{3}\delta_{ij}\text{Tr}(S)(\mathbf{k}) + \text{TF}(S)_{ij}(\mathbf{k}).
\end{equation}
For the purposes of this paper, we are not interested in intrinsic size perturbations, but rather in the intrinsic shape perturbation $g_{ij}(\bb{x})$ and how it correlates with the scalar observable $\delta_D(\bb{x})$ which specifically refers to the galaxy overdensity field. Thus, to be precise, the symmetric 2-tensor field of interest to us is actually 
\begin{equation}\label{eq:stot}
    \bar{S}_{ij}(\mathbf{x}) = \frac{1}{3}\delta_{ij}\delta_D(\bb{x}) + g_{ij}(\bb{x})
\end{equation}
with $g_{ij}$ defined as above. We will henceforth drop the overbar and simply refer to the r.h.s. of the above equation as $S_{ij}(\mathbf{x})$. It is important to realize that our theory model is completely agnostic to such redefinitions. Its general form applies to any tracer of the dark matter field, as we will see below. 

We now concentrate on the large-scale perturbative expansion of the intrinsic shape perturbation $g_{ij}(\bb{x})$ (the bias expansion for scalars such as $\delta_D$ are well-studied elsewhere \cite{schmidt_bias} and we do not reproduce them here, except briefly in Appendix \ref{sec:eft}). The bias expansion for the intrinsic shape perturbation can be separated into two pieces with distinct physical interpretation:
\begin{equation}
\begin{aligned}
    g_{ij}(\bb{x}) &=  g_{ij}^{\text{loc}}(\bb{x}) + g_{ij}^{\text{stoch}}(\bb{x}); \\
    g_{ij}^{\text{loc}}(\bb{x}) &= \sum_\mathcal{O} b^\text{g}_\mathcal{O} \text{TF}(\mathcal{O}_{ij})(\bb{x}); \\
    g_{ij}^{\text{stoch}}(\bb{x}) &= \sum_\mathcal{O} \epsilon_\mathcal{O}(\bb{x}) \text{TF}({O}_{ij})(\bb{x}); 
\end{aligned}
\end{equation}
where the sum is over all independent operators $\mathcal{O}$ that can be measured by a local observer comoving with the tracer. We omit higher-derivative terms here but briefly comment on them at the end of this subsection.

Explicit expressions for each of the terms above can be found in Appendix \ref{sec:eft}. The first term, $g_{ij}^{\text{loc}}(\bb{x})$ corresponds to the local operators (local meaning that they are lowest order in \textit{spatial} derivatives\footnote{A rather unique feature of the EFT of LSS is that it is non-local in time. For this reason, the tracer at some given time $\tau$ depends on the entire history of the dark matter density field in its vicinity. This non-locality in time can be treated perturbatively as well \cite{mirbabayi,leo_bias}. However, at low orders in perturbation theory the nonlocal-in-time operator basis is degenerate with the local-in-time one, and hence this subtlety can be ignored (see also \cite{ansari_nonlocal} for a more recent discussion).}). The authors of \cite{mirbabayi} provide an enumeration of these operators in terms of `building blocks' $\Pi^{[n]}_{ij}$ which can be computed order by order in perturbation theory. 

The second term $g_{ij}^{\text{stoch}}(\bb{x})$ encodes stochastic contributions to the shape perturbation, which is sometimes also referred to as \textit{shape noise}. It corresponds to the influence of short-scale astrophysical processes on the evolution of the shape field on large scales. This can be modelled through the inclusion of stochastic fields $\epsilon(\mathbf{x})$ which have the following properties: (i) they are uncorrelated with the dark matter density on large scales (and therefore uncorrelated with the $\Pi^{[n]}$ operators defined above), (ii) at leading order, they are uncorrelated at different positions, i.e. $\langle \epsilon(\bb{x})\epsilon(\bb{x}')\rangle \propto \delta^D(\mathbf{x-x'}).$ In reality, at any finite separation distance between two objects neither of these statements is exact, since, as shown in great detail in e.g. \cite{schmidt_bias}, it is true that (i) gravitational evolution couples the stochastic fields with the local bias operators $\Pi^{[n]}$, and (ii) the stochastic field value at different positions do have some nonzero correlation due to the spatial nonlocality of tracer formation. All in all, this implies that the deterministic part of the Eulerian bias expansion is supplemented by terms of the form 
\begin{equation}
    g_{ij}^{\text{stoch}}(\bb{x}) \supset \{\epsilon_{ij}(\bb{x}); \epsilon^\delta_{ij}(\bb{x})\delta(\bb{x});\epsilon^\Pi(\bb{x})\text{TF}(\Pi_{ij}^{[1]}(\bb{x})), \dots  \}
\end{equation}
whose correlators schematically satisfy the relation 
\begin{equation}
\begin{aligned}
    \langle& \epsilon_1(\mathbf{x}_1)\epsilon_2(\mathbf{x}_2) \dots\delta(\mathbf{y}_1)\delta(\mathbf{y}_2)\dots\rangle \\
    &= \langle \epsilon_1(\mathbf{x}_1)\epsilon_2(\mathbf{x}_2) \dots\rangle \cdot \langle\delta(\mathbf{y}_1)\delta(\mathbf{y}_2)\dots \rangle;
\end{aligned}  
\end{equation}
that is to say, the stochastic fields $\epsilon_i(\bb{x}_i)$ are uncorrelated with large-scale structure, but their coupling to gravitational evolution is captured by introducing composite operators like $\epsilon^\delta_{ij}(\bb{x})\delta(\bb{x})$. 

Note that these new fields may also carry indices $ij$, in which case they are trace-free. It is expected that the large-scale limit of loop corrections to the bispectrum will naturally require the inclusion of such terms, in similar spirit to the case of galaxy clustering, so that in order to consistently treat the impact of small-scale physics we must introduce these terms with free amplitudes. Put differently, the stochastic terms describe variability in the large-scale bias amplitudes, i.e. the response of the galaxy density field to changes in gravitational observables.  

There is in principle also an additional term $g_{ij}^{\text{h.d.}}(\bb{x})$ which corresponds to higher-derivative corrections coming from the fact that tracer formation is not exactly local in space. There is a characteristic \textit{non-locality scale} $R_*$ which defines the typical size of the region involved in the formation process. These higher-derivative corrections are contractions of partial derivatives $\partial_i$ with the operators $\Pi_{ij}^{[n]}$. Some possible examples would be 
\begin{equation}
    g_{ij}^{\text{h.d.}}(\bb{x}) \supset \{ R_*^2 \nabla^2 \text{TF}(\Pi^{[1]})_{ij}; R_*^2 \text{TF}(\partial_i \partial_k \Pi^{[2]}_{kj}), \dots \}.
\end{equation}
In principle, the scale of spatial non-locality $k_* \sim 1/R_*$ could be different from the nonlinear scale $k_{\text{NL}}$ \cite{fujita}. This depends on the tracer in question, and one can treat each of these scales differently in perturbation theory. In principle, at any order in perturbation theory, all possible contractions of the spatially local operators $\Pi^{[n]}_{ij}$ with derivatives are allowed. The first such term would be of the form 
\begin{equation}\label{eq:hdterm}
    g_{ij}^\text{h.d.}(\bb{x}) = b^\text{g}_{R}(\nabla_i \nabla_j - \frac{1}{3}\nabla^2)\delta(\bb{x}).
\end{equation}
or equivalently in Fourier space 
\begin{equation}
    g_{ij}^\text{h.d.}(\bb{k}) = -b^\text{g}_{R}k^2\mathcal{D}_{ij}(\bb{k})\delta(\bb{k}).
\end{equation}
where $\mathcal{D}_{ij}(\bb{k}) = k^{-2}(\bb{k}_i\bb{k}_j - k^2 \delta_{ij}/3)$. 
For the power spectrum, this leads to familiar corrections $\propto k^2 P_L(k)$. Such terms are strongly degenerate with loop corrections on large scales \cite{bakx_eft}. Our baseline model thus does not include spatially nonlocal terms. 

\subsection{Tree-level Bispectrum}
The bispectrum of shape fields $B_{ijklrs}$ is defined through 
\begin{equation}\label{eq:bispdef}
\begin{aligned}
     \langle S_{ij}(\mathbf{k}_1)S_{kl}(\mathbf{k}_2)S_{rs}(\mathbf{k}_3) \rangle &= (2\pi)^3 \delta^D(\mathbf{k}_1+\mathbf{k}_2+\mathbf{k}_3) \\
     &\times B_{ijklrs}(\mathbf{k}_1,\mathbf{k}_2,\mathbf{k}_3).
\end{aligned}
\end{equation}
The symmetries of the left hand side immediately imply that $B_{ijklrs}$ is symmetric under the exchange of any of the pairs $(ij), (kl), (rs)$ (together with their corresponding wavevector) and also symmetric under exchange of indices within a pair.

After decomposing $S_{ij}$ into a scalar perturbation $\delta$ and an intrinsic shape perturbation $g_{ij}$, the bispectrum naturally decomposes into four different parts: (i) a scalar-scalar-scalar contribution (sss), (ii) a scalar-scalar-shape part (ssg), (iii) a scalar-shape-shape part (sgg) and (iv) a shape-shape-shape part (ggg). For example, 
\begin{align}
    B^{\text{ssg}}_{rs}(\mathbf{k}_1,\mathbf{k}_2,\mathbf{k}_3) &= \langle \text{Tr}(S)(\mathbf{k}_1)\text{Tr}(S)(\mathbf{k}_2)\text{TF}(S_{rs})(\mathbf{k}_3) \rangle ' \nonumber \\
    &= \langle \delta(\mathbf{k}_1)\delta(\mathbf{k}_2)g_{rs}(\mathbf{k}_3) \rangle'.
\end{align}   
Here we use indices $rs$ rather than $ij$ to emphasize that specifically the third field is the tensor variable. Clearly $B^{\text{ssg}}$ is symmetric in its first two arguments $\mathbf{k}_1,\mathbf{k}_2$ while $B^{\text{sgg}}$ is symmetric in the last two arguments $\mathbf{k}_2,\mathbf{k}_3$. The pure-scalar ($B^{\text{sss}}$) and pure-tensor ($B^{\text{ggg}}$) spectra are symmetric in all three wavevectors\footnote{This would not be true if the scalar or shape variables being correlated are different, but we do not consider this scenario here.}. Note that we only need to consider ordered combinations of scalar and shape variables, i.e. we do not need for example shape-scalar-scalar because 
\begin{equation}\label{eq:ordering}
\begin{aligned}
    B^{\text{ssg}}(\mathbf{k}_1,\mathbf{k}_2,\mathbf{k}_3) &= B^{\text{sgs}}(\mathbf{k}_1,\mathbf{k}_3,\mathbf{k}_2) = B^{\text{gss}}(\mathbf{k}_3,\mathbf{k}_1,\mathbf{k}_2); \\
    B^{\text{sgg}}(\mathbf{k}_1,\mathbf{k}_2,\mathbf{k}_3) &= B^{\text{gsg}}(\mathbf{k}_2,\mathbf{k}_1,\mathbf{k}_3) = B^{\text{ggs}}(\mathbf{k}_3,\mathbf{k}_2,\mathbf{k}_1).
\end{aligned}
\end{equation}
Here we suppressed tensor indices to avoid clutter. Since we neglect spatially nonlocal terms in our bias expansion, the bispectrum of shape fields takes the form 
\begin{equation}
    B_{ijklrs}^\text{tree} = B_{ijklrs}^{\text{det,tree}} + B_{ijklrs}^{\text{stoch,tree}}; 
\end{equation}
where the second term contains all contributions with at least one stochastic field. Tree-level contributions to the bispectrum start at fourth order in perturbations and consist of combinations of two first order fields and one second order field. Explicitly\footnote{This expression ignores leading stochastic contributions of the form $\langle \epsilon\epsilon\epsilon\rangle$, but these are listed in Appendix \ref{sec:eft}.},
\begin{equation}\label{eq:wick}
\begin{aligned}
   B_{ijklrs}^{\text{tree}}(\mathbf{k}_1,\mathbf{k}_2,\mathbf{k}_3) &= \langle S_{ij}^{(2)}(\mathbf{k}_1)S_{kl}^{(1)}(\mathbf{k}_2)S_{rs}^{(1)}(\mathbf{k}_3) \rangle' \\
   &+ \langle S_{ij}^{(1)}(\mathbf{k}_1)S_{kl}^{(2)}(\mathbf{k}_2)S_{rs}^{(1)}(\mathbf{k}_3) \rangle'\\
   &+\langle S_{ij}^{(1)}(\mathbf{k}_1)S_{kl}^{(1)}(\mathbf{k}_2)S_{rs}^{(2)}(\mathbf{k}_3) \rangle'. \\
\end{aligned}
\end{equation}
Here the prime denotes a correlator where the factor of $(2\pi)^3 \delta^D(\mathbf{k}_1+\mathbf{k}_2+\mathbf{k}_3)$ ensuring momentum conservation is omitted. The final two terms can be obtained simply by exchanging the index pairs $(ij) \leftrightarrow (kl)$ resp. $(ij) \leftrightarrow (rs)$ together with their corresponding wavevectors. We will thus write this as 
\begin{equation}
    \begin{aligned}
        B_{ijklrs}^{\text{tree}}(\mathbf{k}_1,\mathbf{k}_2,\mathbf{k}_3) = \langle & S_{ij}^{(2)}(\mathbf{k}_1)S_{kl}^{(1)}(\mathbf{k}_2)S_{rs}^{(1)}(\mathbf{k}_3) \rangle' \\
   &+ \text{2 perm.} 
    \end{aligned}
\end{equation} 
The explicit expressions for all the different contributions have mostly been derived elsewhere already \cite{vlah_eft1,schmitz_bisp}. We elaborate on them in Appendix \ref{sec:eft} and also list our definitions of bias coefficients there. We will neglect the effects of infrared resummation on the bispectrum \citep[see e.g.][]{chen_bisp,tspt1,tspt2}, which, although present at the percent level, does not contribute appreciably to the total bispectrum SNR we estimate here. A summary of our theory model can be found in Table \ref{tab:params}. In brief, the total tensor bispectrum of biased tracers depends on seven operator biases $b_\mathcal{O}$ and six stochastic amplitudes. If one only considers bispectra with a single projected tensor field (i.e. E- or B-mode), we only have three stochastic amplitudes, so 10 free parameters in total.

\section{Projected Observables}\label{sec:proj}
An observer at any fixed position in the Universe cannot determine all components of the shape perturbation, because components parallel to the line of sight are not available. Thus, only two of the five degrees of freedom in $g_{ij}$ remain. These are typically labeled E- and B-modes, and they are entirely analogous to the two degrees of freedom of other spin-2 fields on the sky considered previously in the literature, e.g. the cosmic microwave background (CMB) photon polarization \cite{Hu97} and gravitational wave polarization. We now discuss how this projection is modelled for the IA observable. This treatment is identical to that of \cite{vlah_eft2}. Importantly, all findings presented in this Section are completely independent from the theory model from Section \ref{sec:theory} and apply to any LSS observable.

\subsection{Flat Sky Projection}
We can project any trace-free tensorial quantity on the sky by considering the projection operation
\begin{equation}\label{elliptens}
\begin{aligned}
    \gamma_{ij,I}(\mathbf{x}) &:= \text{TF}(\mathcal{P}^{ik}(\hat{\bb{n}})\mathcal{P}^{jl}(\hat{\bb{n}})g_{kl}(\mathbf{x},z)) \\
    &= \mathcal{P}^{ijkl}(\hat{\bb{n}})g_{kl}(\mathbf{x});
\end{aligned}
\end{equation}
where
\begin{equation}
    \mathcal{P}^{ij}(\hat{\bb{n}}):=\delta^{ij}-\hat{\bb{n}}^i\hat{\bb{n}}^j 
\end{equation}
is a projection operator in the $\hat{\bb{n}}$-direction and the last line of Eq. \eqref{elliptens} is the definition of the total projection tensor $\mathcal{P}_{ijkl}$. The projected shape field has two degrees of freedom. We choose to decompose it as 
\begin{equation}\label{eq:projshape}
    \gamma_{ij,I}(\bb{x}) = \bb{M}_{ij}^{(+2)}(\hat{\bb{n}})\gamma_{+2}(\bb{x}) +\bb{M}_{ij}^{(-2)}(\hat{\bb{n}})\gamma_{-2}(\bb{x}) .
\end{equation}
Here $\bb{M}_{ij}^{(\pm 2)}$ are defined through 
\begin{equation}
\begin{aligned}
    \mathbf{M}_{ij} ^{(\pm 2)}(\hat{\bb{n}}) &:= \mathbf{m}_i ^\pm \mathbf{m}_j ^\pm;
\end{aligned}
\end{equation}
where $\mathbf{m}_1 = \hat{\bb{x}}$, $\mathbf{m}_2 = \hat{\bb{y}}$ and $\mathbf{m}^\pm := \mp \frac{1}{\sqrt{2}}(\mathbf{m}_1 \mp \mi\mathbf{m}_2)$. Under complex conjugation we have $(\bb{m}^\pm)^* = -\bb{m}^\mp$ and therefore $\bb{M}_{ij}^{(\pm 2),*}=\bb{M}_{ij}^{(\mp 2)}$. The projection tensors are orthonormal in the complex sense, i.e. $\bb{M}^s\cdot \bb{M}^{*,s'} = \delta_{ss'}$. The components of the shape field in Fourier space are complex valued, but since the configuration-space shape field must be real, we have the constraint $\gamma_{ij,I}^*(\bb{k}) = \gamma_{ij,I}(-\bf{k})$. This enforces the constraint that $\gamma_{\pm 2}^*(\bb{k}) = \gamma_{\mp 2} (-\bb{k})$. In the flat-sky approximation, also referred to as the \textit{global plane-parallel} (GPP) limit, the line of sight is a fixed vector $\hat{\bb{n}}$ and projection operators are position-independent.
Another common basis for this is the E/B-basis also used in analyzing e.g. polarisation of the CMB. This has the advantage that it is independent of the coordinate system chosen (whereas the decomposition in Eq. \eqref{eq:projshape} depends on the choice of specific $x$ and $y$ axes in the plane). It can be obtained via a $\bb{k}$-dependent rotation in the plane of the sky:
\begin{equation}\label{eq:ebdef}
\begin{aligned}
    \gamma_E(\bb{k}) &= \frac{1}{2}(\gamma_{+2}\exp (-2\mi\phi_{\bb{k}})+\gamma_{-2}\exp (+2\mi\phi_{\bb{k}})); \\
    \gamma_B(\bb{k}) &= \frac{1}{2\mi}(\gamma_{+2}\exp(-2\mi\phi_{\bb{k}})-\gamma_{-2}\exp(+2\mi\phi_{\bb{k}}));
\end{aligned} 
\end{equation}
where $\phi_{\bb{k}}$ is the angle the projected $\bb{k}$ vector makes with the (positive) $x$-axis. If we define the E- and B-projection tensors 
\begin{equation}
    \begin{aligned}
        \bb{M}^E_{ij}(\hat{\bb{k}},\hat{\bb{n}}) = \frac{1}{2}\big( & \bb{M}^{(-2)}_{ij}(\hat{\bb{n}})\exp (-2\mi\phi_{\bb{k}}) \\
        &+\bb{M}^{(+2)}_{ij}(\hat{\bb{n}})\exp (+2\mi\phi_{\bb{k}})\big) 
    \end{aligned}
\end{equation}
and
\begin{equation}
    \begin{aligned}
        \bb{M}^B_{ij}(\hat{\bb{k}}, \hat{\bb{n}}) = \frac{1}{2\mi}\big( &\bb{M}^{(-2)}_{ij}(\hat{\bb{n}})\exp (-2\mi\phi_{\bb{k}}) \\
        &-\bb{M}^{(+2)}_{ij}(\hat{\bb{n}})\exp (+2\mi\phi_{\bb{k}})\big),
    \end{aligned}
\end{equation}
then it is clear that 
\begin{equation}
\begin{aligned}
    \gamma_E(\bb{k},\hat{\bb{n}}) &= \bb{M}^E_{ij}(\hat{\bb{k}},\hat{\bb{n}})^*g_{ij}(\bb{k});\\ 
    \gamma_B(\bb{k},\hat{\bb{n}}) &= \bb{M}^B_{ij}(\hat{\bb{k}},\hat{\bb{n}})^*g_{ij}(\bb{k});
\end{aligned}
\end{equation}
and
\begin{equation}
\begin{aligned}
    \bb{M}^E_{ij}(\hat{\bb{k}},\hat{\bb{n}})^* &= \bb{M}^E_{ij}(\hat{\bb{k}},\hat{\bb{n}}); \\ \bb{M}^B_{ij}
    (\hat{\bb{k}},\hat{\bb{n}})^* &= \bb{M}^B_{ij}(\hat{\bb{k}},\hat{\bb{n}});
\end{aligned}
\end{equation}
where we explicitly indicated the $\hat{\bb{k}}$ and $\hat{\bb{n}}$ dependence. The E- and B-fields defined above are real\footnote{Note that $\phi_{-\bb{k}} = \pi + \phi_{\bb{k}}$ so that the phase factor does not change under $\bb{k} \to \bb{-k}$. Indeed for spin-2 fields a rotation by $\pi$ amounts to the identity transformation.} because $\gamma_{E,B}(\mathbf{k})^* = \gamma_{E,B}(-\mathbf{k})$. Also, the E- and B- projection operators are orthogonal and $\bb{M}_X \cdot \bb{M}_X^* = 1/2$. It is important to note that the E- and B-mode projection tensors depend on the wavevector $\bb{k}$ in question. From now on, in order to avoid clutter, we will drop the dependence of the projection tensors and the projected fields on $\hat{\bb{n}}$ (we never consider transformations that change the line of sight direction). 

If we consider a new frame (denoted with a prime) obtained by counterclockwise rotation $\bb{R}_\psi$ around the line of sight by angle $\psi$, the projection tensors transform as spin-weighted 2-tensors in two dimensions, i.e.
\begin{equation}
    \mathbf{M'}^{(\pm 2)}_{i'j'} = (\bb{R}_\psi)^i_{i'}(\bb{R}_\psi)^j_{j'} \mathbf{M} ^{(\pm 2)}_{ij} \exp (\pm 2\mi \psi).
\end{equation}
The angle $\phi_{\bb{k}}$ transforms as $\phi'_{\bb{k}'} = \phi_{\bb{k}} - \psi$. It is hence clear that $\gamma_{E,B}'(\bb{k}') = \gamma_{E,B}(\bb{k})$. It is convenient to also introduce a `scalar projection operator' for the scalar tracer $D$ (c.f. Eq. \ref{eq:stot}) in the trace of $S_{ij}$;
\begin{equation}
    \bb{M}_{ij}^D(\hat{\bb{k}}) = \delta_{ij}
\end{equation}
so that we have 
\begin{equation}
    X(\bb{k}) = \bb{M}_{ij}^X(\hat{\bb{k}})S_{ij}(\bb{k})
\end{equation}
for $X=D,E,B$\footnote{Note the slight abuse of notation - it would be more precise to write $\delta_D$ instead of $D$ and $\gamma_{E,B}$ instead of $E,B$.}). Then we can define the following \textit{coordinate-independent} bispectra: 
\begin{equation}\label{eq:indep}
    B_{XYZ} = \bb{M}_{ij}^{X}(\hat{\bb{k}}_1) \bb{M}_{kl}^{Y}(\hat{\bb{k}}_2)\bb{M}_{rs}^{Z}(\hat{\bb{k}}_3)B_{ijklrs};  
\end{equation}\
where $X,Y,Z= D,E,B$ and we omit the arguments $(\mathbf{k}_1,\mathbf{k}_2,\mathbf{k}_3)$ on both sides. If we consider a \textit{reflection} $\bb{P}$ that keeps the line of sight and the $y$-axis fixed but sends $\hat{\bb{x}}$ to $-\hat{\bb{x}}$, then we get $\phi'_{\bb{k}'} = \pi - \phi_{\bb{k}}$ and $\bb{m'}^\pm = \bb{m}^\mp$. Thus,
\begin{equation}\label{eq:ebrefl}
\begin{aligned}
     (\bb{M}^E_{i'j'})(\hat{\bb{k}}') &= \bb{P}^i_{i'}\bb{P}^j_{j'}\bb{M}^E_{ij}(\hat{\bb{k}}), \\
     (\bb{M}^B_{i'j'})(\hat{\bb{k}}') &= -\bb{P}^i_{i'}\bb{P}^j_{j'}\bb{M}^B_{ij}(\hat{\bb{k}}),
\end{aligned}
\end{equation}
while $\bb{M}_{ij}^D$ obviously remains invariant. Note that these transformation laws are completely general, i.e. they do not assume anything about statistical isotropy or parity invariance of the resulting spectra. 

In contrast to the power spectrum case, parity conservation does \textit{not} impose that bispectra with an odd number of B-modes vanish \cite{schmitz_bisp}. As such, in general all possible combinations of $D, E, B$ constitute nontrivial bispectra even in absence of parity-violating physics, i.e. we have the 10 possible combinations 
\begin{equation}\label{eq:combs}
    \begin{aligned}
        \text{sss}&: B_{DDD}; \\
        \text{ssg}&: B_{DDE}, B_{DDB}; \\
        \text{sgg}&: B_{DEE}, B_{DEB}, B_{DBB}; \\
        \text{ggg}&: B_{EEE}, B_{EEB}, B_{EBB}, B_{BBB}. \\
    \end{aligned}
\end{equation}
This appears to contradict earlier findings from \cite{schneider_parity}. We elaborate on this issue in Appendix \ref{sec:parity}. Importantly, Eq. \eqref{eq:ebrefl} implies that when considering tensor bispectra, those with an odd number of B-modes are \textit{odd} under parity transformations \footnote{In this work, we do not concern ourselves with any parity-violating phenomena. We simply want to express the fact that bispectra with an odd number of B-modes possess `handedness', i.e. if the coordinate system in the sky plane is reflected, they change sign. In that sense, it is appropriate to call them `parity-odd'. }, i.e.
\begin{equation}
    B_{XYZ}(\bb{P}\bb{k}_1,\bb{P}\bb{k}_2,\bb{P}\bb{k}_3) = -B_{XYZ}(\bb{k}_1,\bb{k}_2,\bb{k}_3)
\end{equation}
if an odd number of $X,Y,Z$ is a B-mode.
This will mean that their multipole expansion is different from parity-even multipoles.

Analogous to the notation above we can form kernels for scalar ($D$), E- and B-modes up to first and second order, i.e. $\mathcal{K}_{X}^{(1)}(\mathbf{k})$ and $\mathcal{K}_{X}^{(2)}(\mathbf{k}_1,\mathbf{k}_2)$ for $X = D,E,B$ (see Appendix \ref{sec:eft}). There are further simplifications at tree-level. There, exactly two of the three fields entering the bispectrum must be of first order. However, since the first order E- and B-mode kernels of the shape field $g_{ij}$ satisfy 
\begin{equation}
\begin{aligned}
    \mathcal{K}_E^{(1)}(\mathbf{k}) &= \bb{M}^E_{ij}(\hat{\bb{k}}) \cdot \mathcal{K}_{ij}^{\text{g},(1)}(\mathbf{k}) = \frac{1}{2}(1-\mu_\bb{k}^2); \\
    \mathcal{K}_B^{(1)}(\mathbf{k}) &= \bb{M}^B_{ij}(\hat{\bb{k}}) \cdot \mathcal{K}_{ij}^{\text{g},(1)}(\mathbf{k}) = 0;
\end{aligned}  
\end{equation}
it follows that the deterministic parts of spectra with at least two B-fields must be zero at tree-level. 

\subsection{Bispectrum Multipoles}
\begin{figure}
    \centering
    \includegraphics[width=\linewidth]{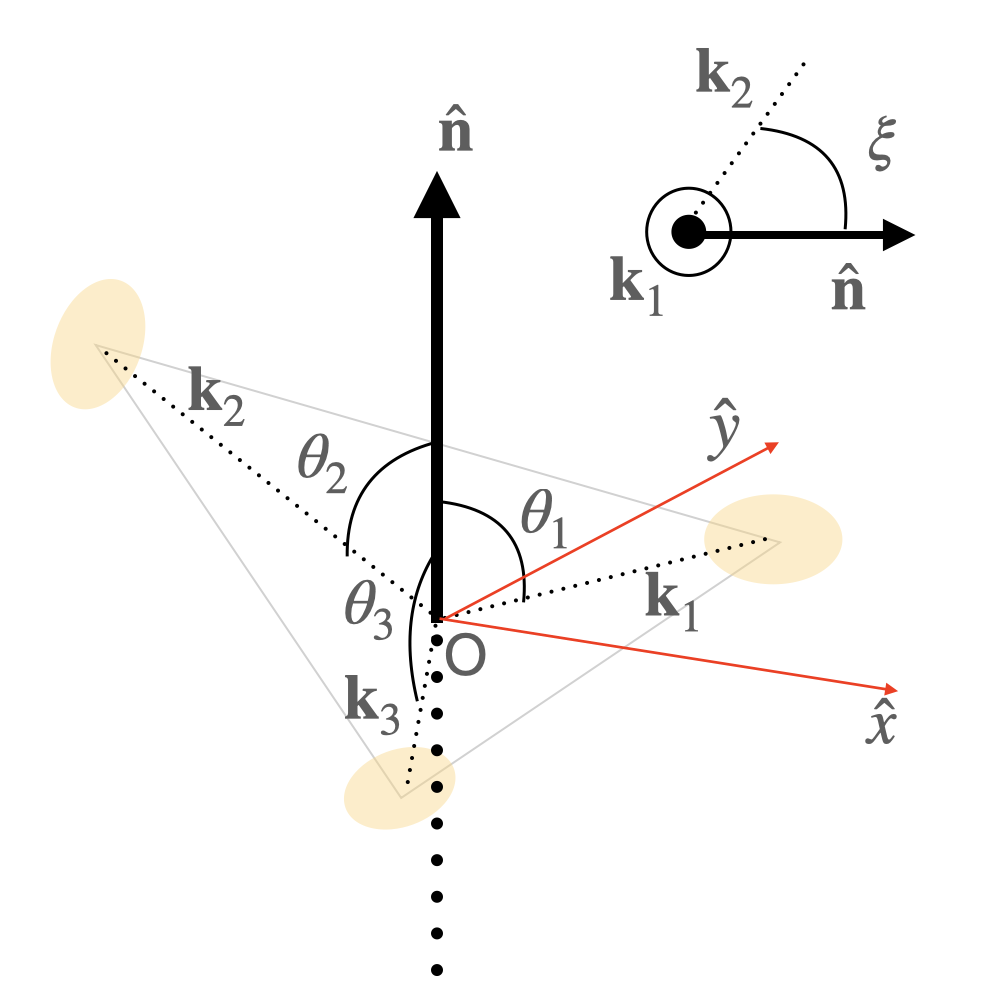}
    \caption{A graphical illustration of the geometry of the IA bispectrum. The galaxy shapes (beige) are projected on the $xy$ plane (recall that we take $\hat{\bb{n}}$ to point along the $z$-axis), leading to a non-trivial dependence on the angles $\theta_i$ (where $\mu_i = \cos \theta_i$) already in the absence of redshift-space distortions. The inset in the top right illustrates the definition of $\xi$; it is the angle between $\bb{k}_2$ and $\hat{\bb{n}}$ when viewed along $\bb{k}_1$.}
    \label{fig:bispgeom}
\end{figure}

Since the bispectra defined in Eq. \eqref{eq:indep} are manifestly coordinate invariant, we can, in analogy to the literature on the redshift-space bispectrum \cite{scoccimarro_couchman} express them in terms of the five variables\footnote{Other angular bases have also been considered in e.g. \cite{hashimoto_bisp,gualdi_bisp,gualdi_bisp_trisp} and these would also be suitable.}
\footnote{The angle $\xi$ is undefined for collinear triangles, i.e. when all wavevectors are collinear. See \cite{oddo_bisp,biagetti_cov} for an extended discussion on open or collinear triangle bins.} $k_1,k_2,k_3,\mu_1, \xi$. Here, the $k_i$ are the side lengths of the triangle, $\mu_i = \hat{\bb{k}}_i \cdot \hat{\bb{n}}$ and $\xi$ describes an azimuthal rotation of $\bb{k}_2$ around $\bb{k}_1$ (where $\xi=0$ corresponds to the case where $\bb{k}_2$ is `in between' $\bb{k}_1$ and $\hat{\bb{n}}$) \cite{scoccimarro_couchman}. We illustrate the geometry in Figure \ref{fig:bispgeom}.
Explicitly, we can put $\bb{k}_1$ in the $xz$-plane (with $\bb{k}_{1,x} >0$) without loss of generality, in which case we have  
\begin{equation}\label{eq:unitvecs}
    \begin{aligned}
        \hat{\bb{k}}_1 &= (\sqrt{1-\mu_1^2},0,\mu_1); \\
        \hat{\bb{k}}_2 &= (\mu_1\sqrt{1-\mu_{12}^2}\cos \xi + \mu_{12}\sqrt{1-\mu_1^2},\sqrt{1-\mu_{12}^2}\sin \xi, \\
        &\mu_1\mu_{12}-\sqrt{1-\mu_{12}^2}\sqrt{1-\mu_{1}^2}\cos\xi); \\
        \bb{k}_3 &= -\bb{k}_1-\bb{k}_2.
    \end{aligned}
\end{equation}
The resulting expressions for the tree-level bispectra are rather lengthy functions of the five variables $k_1,k_2,k_3,\mu_1$ and $\xi$, which we do not reproduce here (but see Appendix \ref{sec:eft}). 

It is easy to see that flipping $\xi$ to $-\xi$ amounts to performing a reflection in the $xz$-plane. Thus, parity-odd bispectra are \textit{odd} functions of the angular variable $\xi$. Conversely, parity-even bispectra are even functions of $\xi$. This will have important ramifications when we construct bispectrum multipole estimators.  

Rather than retaining the full angular dependence of the bispectrum, it is common to consider its multipole moments as a means of further compressing the bispectrum data vector. In the most general sense, multipole moments should be viewed as spherical averages of the full 3D bispectrum over all possible triangle orientations with some angular weighting, i.e. schematically we can write
\begin{equation}\label{eq:intmult}
\begin{aligned}
    B^L_{XYZ}(k_1,&k_2,k_3) = \frac{N_L}{V_\Delta} \int_{k_{1,2,3}} \md^3\bb{q}_{1,2,3} \,\delta^D(\bb{q}_{123}) \\ &\times \bigg(\mathcal{P}_{XYZ}^{L}(\{\hat{\bb{q}}_i\}) B_{XYZ}(\mathbf{q}_1,\mathbf{q}_2,\mathbf{q}_3)\bigg).
\end{aligned}
\end{equation}
where $N_L$ is a normalization depending on the multipole and the notation for the integration region is 
\begin{equation}
    \int_{k_{1,2,3}} \md^3\bb{q}_{1,2,3} = \prod_{i=1,2,3}\bigg(\int_{k_i-\Delta k/2}^{k_i+\Delta k/2}q_i^2 \md q_i \int \md \Omega_i\bigg).
\end{equation}
and $V_\Delta$ is the volume of this region. Thus, the multipole moments only depend on the side lengths. The angular weighting $\mathcal{P}_{XYZ}^{L}(\{\hat{\bb{q}}_i\})$ can depend on the fields $XYZ$ considered, and the triple solid angle integration is over all orientations of the unit vectors. Since the bispectra involving E- and B-modes are invariant under 2D rotations in the sky plane, they now only depend on the side lengths of the triangle and the orientation of the triangle with respect to the line of sight. Furthermore, we can take the radial integral to be trivial if the bins are small enough\footnote{In principle, the estimator for the bispectrum multipoles is a discrete average over all wavevectors that enter a given triangle bin. Throughout this entire paper we neglect differences between this discrete average and the integral approximation from Eq. \eqref{eq:intmult}. We address this issue in more detail in our companion paper \cite{paper2}.}. For these reasons, one can use the angular variables $\mu_1,\xi$ as mentioned above and convert the triple momentum integration into an integral over these two angular variables \cite{oddo_bisp,rizzo_bisp}:
\begin{equation}
    \frac{1}{V_\Delta}\int_{k_{1,2,3}} \md^3\bb{q}_{1,2,3}\,\delta^D(\mathbf{q}_{123}) \approx \frac{1}{4\pi}\int \md \mu_1 \md \xi.
\end{equation}
It is always understood that the integration bounds for $\mu_1$ resp. $\xi$ are $[-1,1]$ and $[0,2\pi]$. Alternatively, this angular average is sometimes viewed as an average over all orientations of the line of sight, if the triangle is assumed to be in a fixed position. Thus, in full analogy to the galaxy bispectrum multipoles in redshift space, one can consider a complete expansion of the form 
\begin{equation}
    \begin{aligned}
        B_{XYZ}(k_1,k_2,k_3,\mu_1,\xi) &= \sum_{\ell,m}\bigg(Y_{\ell m}(\mu_1,\xi)\\
        &\times B_{XYZ}^{\ell m}(k_1,k_2,k_3)\bigg)
    \end{aligned}
\end{equation}
where the expansion coefficients are given by 
\begin{equation}\label{eq:expcoeff}
    \begin{aligned}
        B_{XYZ}^{\ell m}(k_1,k_2,k_3) &= \frac{2\ell+1}{4\pi}\int \md \mu_1 \md \xi \bigg(Y_{\ell m}(\mu_1,\xi)\\ 
        &\times B_{XYZ}(k_1,k_2,k_3,\mu_1,\xi)\bigg)
    \end{aligned}
\end{equation}
and the prefactor is due to the normalization of the spherical harmonics. We choose to work with \textit{real} spherical harmonics $Y_{\ell m}$ (denoted with two lower indices, see also \cite{nann}), which are defined through a simple basis transformation
\begin{equation}
\begin{aligned}
        Y_{\ell(m<0)} &= \frac{\mi}{\sqrt{2}} (Y_\ell^m - (-1)^m Y_\ell^{-m});\\
        Y_{\ell(m=0)} &= Y_{\ell}^{0}; \\ 
        Y_{\ell(m>0)}&= \frac{1}{\sqrt{2}} (Y_\ell^m + (-1)^m Y_\ell^{-m}).
\end{aligned}
\end{equation}
A basis of real spherical harmonics ensures that the resulting multipoles are real, since the bispectrum is itself real. Here $Y_\ell^m$ (with one lower index) is the usual spherical harmonic (see Eq. \eqref{eq:sphdef}). Spherical harmonics with $m\geq 0$ will be referred to as \textit{cosine type} and those with $m<0$ as \textit{sine type}, since they are proportional to $\cos m\xi$ and $\sin m\xi$ respectively. Parity-even (odd) bispectra can thus be entirely decomposed into spherical harmonics of cosine (sine) type. Concretely, we have the two disjoint expansions 
\begin{equation}\label{eq:evenmult}
    \begin{aligned}
        B_{XYZ}(k_1,k_2,k_3,\mu_1,\xi)&_\text{even} = \sum_{\ell,m\geq 0}\bigg(Y_{\ell m}(\mu_1,\xi)\\
        &\times B_{XYZ}^{\ell m}(k_1,k_2,k_3)\bigg)
    \end{aligned}
\end{equation}
and 
\begin{equation}\label{eq:oddmult}
    \begin{aligned}
        B_{XYZ}(k_1,k_2,k_3,\mu_1,\xi)&_\text{odd} = \sum_{\ell,m<0}\bigg(Y_{\ell m}(\mu_1,\xi)\\
        &\times B_{XYZ}^{\ell m}(k_1,k_2,k_3)\bigg).
    \end{aligned}
\end{equation}
If we consider the case where $X,Y,Z$ are all scalars (so that $XYZ$ is even), then as we prove in Appendix \ref{sec:angbasis} we can equivalently define the set of multipoles
\begin{equation}\label{eq:newbasis}
\begin{aligned}
    B^{O,\ell_1\ell_2}_{\text{sss}}(k_1,k_2,k_3) &= \frac{N_{\ell_1\ell_2}}{4\pi}\int\md\mu_1 \md\xi \bigg(\mathcal{L}_{\ell_1}(\mu_1)\mathcal{L}_{\ell_2}(\mu_2) \\
    &\times B_{\text{sss}}(k_1,k_2,k_3,\mu_1,\xi) \bigg) 
\end{aligned}
\end{equation}
with the (arbitrary) normalization
\begin{equation}\label{eq:nml}
    N_{\ell_1\ell_2} = (2\ell_1+1)(2\ell_2+1)
\end{equation}
where $\ell_1,\ell_2 \geq 0$ and $\mathcal{L}_\ell$ an ordinary Legendre polynomial (hence the superscript $O$). See Eq. \eqref{eq:mu2mu3rel} for the relation between $\mu_{2,3}$ and $\mu_1,\xi$. Thus, every multipole $B^{\ell m}$ from Eq. \eqref{eq:expcoeff} is a linear combination of the $B^{\ell_1\ell_2}$ from above and vice versa. However, Eq. \eqref{eq:newbasis} is in general more convenient because it allows for a straightforward separable estimator using FFTs. It is also a more suitable starting point for building more general estimators which involve tensor fields. 

The arguments from Appendix \ref{sec:angbasis} also show more generally that parity-even tensor multipoles can be equivalently defined in a double ordinary Legendre basis:
\begin{equation}\label{eq:newbasis_even}
\begin{aligned}
    B^{O,\ell_1\ell_2}_{XYZ}(k_1,k_2,k_3)_\text{even} &= \frac{N_{\ell_1\ell_2}}{4\pi}\int\md\mu_1 \md\xi \bigg(\mathcal{L}_{\ell_1}(\mu_1)\mathcal{L}_{\ell_2}(\mu_2) \\
    &\times B_{XYZ}(k_1,k_2,k_3,\mu_1,\xi)_\text{even} \bigg). 
\end{aligned}
\end{equation}
while the parity-odd tensor multipoles can be written in a similar basis: 
\begin{equation}\label{eq:newbasis_odd}
\begin{aligned}
    B & ^{O,\ell_1\ell_2}_{XYZ}(k_1, k_2,k_3)_\text{odd} =-\frac{N_{\ell_1\ell_2}}{4\pi}\int\md\mu_1 \md\xi \bigg(\mathcal{L}_{\ell_1}(\mu_1)\mathcal{L}_{\ell_2}(\mu_2) \\
    &\times \sqrt{\frac{4\pi}{3}(1-\mu_{12}^2)} Y_{1\,-1}(\mu_1,\xi) B_{XYZ}(k_1,k_2,k_3,\mu_1,\xi)_\text{odd} \bigg) 
\end{aligned}
\end{equation}
where $Y_{1\,-1}(\mu_1,\xi) = -\sqrt{\frac{3}{4\pi}}\sqrt{1-\mu_1^2}\sin \xi$ is the first parity-odd spherical harmonic. By Eq. \eqref{eq:unitvecs},
\begin{equation}\label{eq:cross2}
    -\sqrt{\frac{4\pi}{3}(1-\mu_{12}^2)} Y_{1\,-1}(\mu_1,\xi) = \hat{\bb{n}} \cdot (\hat{\bb{k}}_1 \times \hat{\bb{k}}_2)
\end{equation}
so that this expression is well-defined even for collinear triangles\footnote{The overall minus sign in this expression is the reason for the overall minus sign in Eq. \eqref{eq:newbasis_odd}. Indeed, the projectors defined in Eqs. \eqref{eq:ordproject},\eqref{eq:asproject} do not contain any overall minus signs, and neither do the estimators.}. 

As we will see, it is possible to construct separable Fourier-space estimators for all tensor multipoles $B^{O,\ell_1\ell_2}_{XYZ}$ \textit{in the global plane-parallel limit} that can be computed via FFTs. We will refer to these definitions as \textit{ordinary Legendre multipoles} or OLM's. We can write them more succinctly by introducing even and odd projectors $\mathcal{P}_{XYZ}^{O,\ell_1\ell_2}(\hat{\bb{k}}_1,\hat{\bb{k}}_2,\hat{\bb{k}}_3)$, which are given by 
\begin{equation}\label{eq:ordproject}
\begin{aligned}
     \mathcal{P}_{XYZ}^{O,\ell_1\ell_2}(\{\hat{\bb{k}}_i\})_\text{even} &= 
     \mathcal{L}_{\ell_1}(\mu_1)\mathcal{L}_{\ell_2}(\mu_2);\\
     \mathcal{P}_{XYZ}^{O,\ell_1\ell_2}(\{\hat{\bb{k}}_i\})_\text{odd} &=
     \mathcal{L}_{\ell_1}(\mu_1)\mathcal{L}_{\ell_2}(\mu_2)\\
     &\times \big(\hat{\bb{n}} \cdot (\hat{\bb{k}}_1 \times \hat{\bb{k}}_2)\big).
\end{aligned}
\end{equation}
These depend on the orientation of the triangle with respect to the line of sight (hence the notation involving the set of three vectors $\{\bb{k}_i\}$ with $i=1,2,3$). 
From now on we omit the `even' and `odd' labels of the projectors and the corresponding bispectra, since this is already implicitly determined from the choice of $XYZ$. 
The OLMs for the bispectrum (that is, Eqs. \ref{eq:newbasis_even},\ref{eq:newbasis_odd}) are thus compactly written as 
\begin{equation}\label{eq:olmdef}
\begin{aligned}
    B^{O,\ell_1\ell_2}_{XYZ}(k_1,k_2,k_3) &= \frac{N_{\ell_1\ell_2}}{4\pi}\int\md\mu_1 \md\xi \bigg(\mathcal{P}_{XYZ}^{O,\ell_1\ell_2}(\{\hat{\bb{k}}_i\}) \\
    &\times B_{XYZ}(k_1,k_2,k_3,\mu_1,\xi) \bigg) 
\end{aligned}
\end{equation}
with $\ell_1,\ell_2 \geq 0$ and $N_{\ell_1\ell_2}$ as in Eq. \eqref{eq:nml}. When $X,Y,Z$ are all scalars (and thus $XYZ$ is necessarily even), this definition reduces to the one from Eq. \eqref{eq:newbasis}; thus, this is a tensorial generalization of the scalar multipole definition. 

\subsection{Associated projectors}
We will also consider an alternative definition which uses \textit{associated} Legendre polynomials for the angular weighting. The need for this modification was considered in \cite{kurita_window}, and we return to it below. In brief, the associated Legendre polynomials serve the purpose of multiplying the $E$ and $B$ fields by an additional factor of $1-\mu_i^2$ before expanding in the two angles $\mu_1$ and $\mu_2$. This is done in order to cancel the denominator in the phase factor from Eq. \eqref{eq:ebdef}, viz. 
\begin{equation}
    \exp (\pm 2 \mi\phi_{a}) = \frac{(\bb{k}_{a,x} \pm \mi \bb{k}_{a,y})^2}{1-\mu_a^2}.
\end{equation}
where $a=1,2,3$. Cancellation of this denominator is not strictly necessary for our local plane-parallel estimators, but becomes more important for global plane-parallel estimators (see Sect. \ref{sec:gpp}). More precisely, we consider the even and odd `associated projectors', now with a superscript $A$, 
\begin{equation}\label{eq:asproject}
\begin{aligned}
     \mathcal{P}_{XYZ}^{A,\ell_1\ell_2}(\{\hat{\bb{k}}_i\})_\text{even} &= 
     \mathcal{L}^{m_X}_{\ell_1}(\mu_1)\mathcal{L}^{m_Y}_{\ell_2}(\mu_2)\mathcal{L}^{m_Z}_{m_Z}(\mu_3);\\
     \mathcal{P}_{XYZ}^{A,\ell_1\ell_2}(\{\hat{\bb{k}}_i\})_\text{odd} &=
     \mathcal{L}^{m_X}_{\ell_1}(\mu_1)\mathcal{L}^{m_Y}_{\ell_2}(\mu_2)\mathcal{L}^{m_Z}_{m_Z}(\mu_3)\\
     &\times \big(\hat{\bb{n}} \cdot (\hat{\bb{k}}_1 \times \hat{\bb{k}}_2)\big).
\end{aligned}
\end{equation}
The indices $m_{X,Y,Z}$ are equal to the spin of the respective field, i.e. $m_X = 0\,(2)$ if $X$ is a scalar (tensor). Finally, $\mathcal{L}_\ell^m$ ($\ell\geq m$) is an associated Legendre polynomial whose definition is given in Eq. \eqref{eq:legdef}. The last factor $\mathcal{L}^{m_Z}_{m_Z}$ is the same for any multipole, and simply equals $1$ if $m_Z = 0$ and $3(1-\mu_3^2)$ if $m_Z = 2$. The \textit{associated Legendre multipoles} or ALMs for the bispectrum are then given as 
\begin{equation}\label{eq:almdef}
\begin{aligned}
    B^{A,\ell_1\ell_2}_{XYZ}(k_1,k_2,k_3) &= \frac{N_{\ell_1\ell_2}}{4\pi}\int\md\mu_1 \md\xi \bigg(\mathcal{P}_{XYZ}^{A,\ell_1\ell_2}(\{\hat{\bb{k}}_i\}) \\
    &\times B_{XYZ}(k_1,k_2,k_3,\mu_1,\xi) \bigg) 
\end{aligned}
\end{equation}
Strictly speaking, the full set of OLMs and ALMs contain the same amount of information. However, in practice only finitely many multipoles will be used and there may be observable differences between the two \cite{inoue}. Multiplication by $1-\mu^2$ effectively downweights radial contributions with $\mu^2 \sim 1$ and may therefore have a discernible impact on the SNR; we will examine this in more detail in Section \ref{sec:results}, but as we will see the overall impact is modest.
\subsection{Global and Local Plane-Parallel Estimators}\label{sec:gpp}
As was the case for Eq. \eqref{eq:newbasis}, there exist simple separable estimators for both OLMs and ALMs in the global plane-parallel (GPP) or flat-sky limit. They can be defined via \cite{rizzo_bisp}
\begin{equation}
\begin{aligned}
    \hat{B}^{T,\ell_1\ell_2}_{XYZ}(& k_1,k_2,k_3) = \frac{N_{\ell_1\ell_2}}{N_\Delta V}\sum_{\mathbf{q}_i \in \text{bin } i}\delta^K(\mathbf{q}_{123}) \\
    &\times \hat{X}(\mathbf{q}_1) \hat{Y}(\mathbf{q}_2)\hat{Z} (\mathbf{q}_3)\mathcal{P}_{XYZ}^{T,\ell_1\ell_2}(\{\hat{\bb{q}}_i\})
\end{aligned}
\end{equation}
where the type $T$ is either $O$ or $A$, $N_\Delta$ is the number of triangles in the bin and $V = L^3$ is the volume of the simulation box. Hats on the fields indicate that these are measured quantities. In the even case, this estimator is given in separable form via 
\begin{equation}\label{eq:sepeq}
    \begin{aligned}
        \hat{B}_{XYZ}^{T,\ell_1\ell_2}&(k_1,k_2,k_3)_{\text{even}} =\frac{N_{\ell_1\ell_2}}{N_\Delta V}\sum_{\mathbf{q}_i \in \text{bin } i}\delta^K(\mathbf{q}_{123})\\
    &\times \tilde{X}^{m_X}_{\ell_1}(\bb{q}_1)\tilde{Y}^{m_Y}_{\ell_2}(\bb{q}_2)\tilde{Z}^{m_Z}(\bb{q}_3)
    \end{aligned}
\end{equation}
where the auxiliary fields are 
\begin{equation}\label{eq:evenest}
\begin{aligned}
    \tilde{X}^{m_X}_{\ell_1}(\bb{q}_1) &= \hat{X}(\mathbf{q}_1) \times \big( \mathcal{L}_{\ell_1}(\mu_1) \text{ or }\mathcal{L}^{m_X}_{\ell_1}(\mu_1)\big) ; \\
    \tilde{Y}^{m_Y}_{\ell_2}(\bb{q}_2) &= \hat{Y}(\mathbf{q}_2) \times \big( \mathcal{L}_{\ell_2}(\mu_2) \text{ or } \mathcal{L}^{m_Y}_{\ell_2}(\mu_2)\big); \\
    \tilde{Z}^{m_Z}(\bb{q}_3) &= \hat{Z}(\mathbf{q}_3) \times \big(1 \text{ or } \mathcal{L}^{m_Z}_{m_Z}(\mu_3)\big);
\end{aligned}
\end{equation}
depending on whether $T=O$ or $A$. It can be computed using a single FFT \footnote{Observe however that some auxiliary fields may be purely imaginary, i.e. $\tilde{X}(\bb{q})^* = - \tilde{X}(-\bb{q})$.}. In the $XYZ$ odd case, we can use Eq. \eqref{eq:cross2} to get 
\begin{equation}\label{eq:oddest}
    \begin{aligned}
        \hat{B}_{XYZ}^{T,\ell_1\ell_2}&(k_1,k_2,k_3)_{\text{odd}} =\frac{N_{\ell_1\ell_2}}{N_\Delta V}\sum_{\mathbf{q}_i \in \text{bin } i}\delta^K(\mathbf{q}_{123})\\
    &\times \bigg(\tilde{X}^{m_X}_{\ell_1,x}(\bb{q}_1)\tilde{Y}^{m_Y}_{\ell_2,y}(\bb{q}_2)\tilde{Z}^{m_Z}(\bb{q}_3) - (x \leftrightarrow y)\bigg)
    \end{aligned}
\end{equation}
where the auxiliary fields are 
\begin{equation}
\begin{aligned}
    \tilde{X}^{m_X}_{\ell_1,a}(\bb{q}_1) &= \hat{X}(\mathbf{q}_1) \hat{\bb{q}}_{1,a} \times \big( \mathcal{L}_{\ell_1}(\mu_1) \text{ or }\mathcal{L}^{m_X}_{\ell_1}(\mu_1)\big);\\
    \tilde{Y}^{m_Y}_{\ell_2,a}(\bb{q}_2) &= \hat{Y}(\mathbf{q}_2) \hat{\bb{q}}_{2,a} \times \big( \mathcal{L}_{\ell_2}(\mu_2) \text{ or } \mathcal{L}^{m_Y}_{\ell_2}(\mu_2)\big);\\
    \tilde{Z}^{m_Z}(\bb{q}_3) &= \hat{Z}(\mathbf{q}_3) \times \big(1 \text{ or } \mathcal{L}^{m_Z}_{m_Z}(\mu_3)\big)
\end{aligned}
\end{equation}
for $a=x,y$. Note that Eq.~\eqref{eq:oddest} requires two bispectrum measurements; it is a sum-separable expression. 
Hence, in the global plane-parallel limit, there is no difference in computational complexity between OLMs and ALMs. However, following \cite{kurita_window}, from the perspective of actual observational data analysis there are known advantages of using associated Legendre polynomials as a basis for shape fields.
The redshift distortion of the observed galaxy density fluctuations and the projection of galaxy shapes onto the sky (Eq.~\ref{eq:projshape}) are defined with respect to the line of sight of each galaxy. 
In realistic wide-angle galaxy surveys, it is not possible to assume a uniform line of sight for all galaxies. 
In such cases, the \textit{local plane-parallel} (LPP) estimator, also called the Yamamoto estimator \cite{yamamoto}, has been employed to correctly measure the anisotropy of statistics with respect to the local line of sight. 
The estimator for the galaxy (density) bispectrum was proposed in  \cite{scoccimarro_fast} \citep[Eq.~(31) of][]{scoccimarro_fast}. 
Sticking to the parity-even case for the sake of exposition, this would be extended to the bispectrum estimator for three arbitrary observed fields $\hat{X}$, $\hat{Y}$, and $\hat{Z}$ of interest, as 
\begin{align}\label{eq:yamamoto}
    \hat{B}_{XYZ}^{A,\ell_1 \ell_2}(&k_1,k_2,k_3) 
    \equiv 
    \frac{N_{\ell_1 \ell_2}}{N_\Delta} 
    \int_{k_{1,2,3}}\mathrm{d}^3\bb{q}_{1,2,3} \delta^D(\mathbf{q}_{123}) \nonumber\\
    &\times \left( \int_{\mathbf{x}_1} \hat{X}(\mathbf{x}_1) e^{-\mi\mathbf{q}_1\cdot\mathbf{x}_1} 
    \mathcal{L}_{\ell_1}^{m_X}(\hat{\mathbf{q}}_1\cdot\hat{\mathbf{d}}) \right) \nonumber\\
    &\times \left( \int_{\mathbf{x}_2} \hat{Y}(\mathbf{x}_2) e^{-\mi\mathbf{q}_2\cdot\mathbf{x}_2} \mathcal{L}_{\ell_2}^{m_Y}(\hat{\mathbf{q}}_2\cdot\hat{\mathbf{d}}) \right) \nonumber\\
    &\times \left( \int_{\mathbf{x}_3} \hat{Z}(\mathbf{x}_3) e^{-\mi\mathbf{q}_3\cdot\mathbf{x}_3}  \mathcal{L}_{m_Z}^{m_Z}(\hat{\mathbf{q}}_3\cdot\hat{\mathbf{d}}) \right), 
\end{align}
where $\hat{\mathbf{d}}$ is a unit vector towards the local region where the galaxy triplet ($\mathbf{x}_1, \mathbf{x}_2,\mathbf{x}_3$) is located. 
The (auxiliary) local E-mode field, setting $\hat{Z}\equiv\hat{E}$ for instance, is defined from the observed projected shape field $\hat{\gamma}_{\pm2}(\mathbf{x}_3)$ with $\mathbf{q}_3$-dependent rotations on the plane perpendicular to the local line of sight $\hat{\mathbf{d}}$ as 
\begin{align}
    \hat{E}(\mathbf{x}_3) &\equiv \frac{1}{2} \left( \hat{\gamma}_{+2}(\mathbf{x}_3)e^{-2\mi\phi(\mathbf{q}_3,\mathbf{d})} + \hat{\gamma}_{-2}(\mathbf{x}_3)e^{+2\mi\phi(\mathbf{q}_3,\mathbf{d})} \right) 
\end{align}
where 
\begin{align}\label{eq:phasefac}
    e^{\pm2\mi\phi(\mathbf{q}_3,\mathbf{d})} \equiv \frac{\mathbf{M}_{ij}^{(\pm2)}(\hat{\mathbf{d}})\hat{\bb{q}}_{3,i}\hat{\bb{q}}_{3,j}}{1-(\hat{\mathbf{q}}_3\cdot\hat{\mathbf{d}})^2}. 
\end{align}
The B-mode can be defined in a similar way. 
In practice, we further employ the endpoint approximation, e.g., $\hat{\mathbf{d}}\simeq\hat{\mathbf{x}}_1$, which allows us to estimate the LPP bispectrum using a small number of FFTs due to the sum-separability of polynomials of cosines, $\hat{\mathbf{q}}\cdot\hat{\mathbf{x}}=\hat{\bb{q}}_i\hat{\bb{x}}_i$ \cite{scoccimarro_fast,bianchi} (see also \cite{philcox_beyond} for a generalization). 
Notice that the phase factor in the definition of the above local E-mode field, by itself, involves a division of polynomials of the cosine, making it non-separable. To resolve this problem, one uses associated Legendre polynomials ($m_Z=2$) as a basis rather than ordinary Legendre polynomials ($m_Z=0$), which can cancel out the denominator \cite{kurita_window}. 
\begin{figure*}[t]
    \centering
    \includegraphics[width=\linewidth]{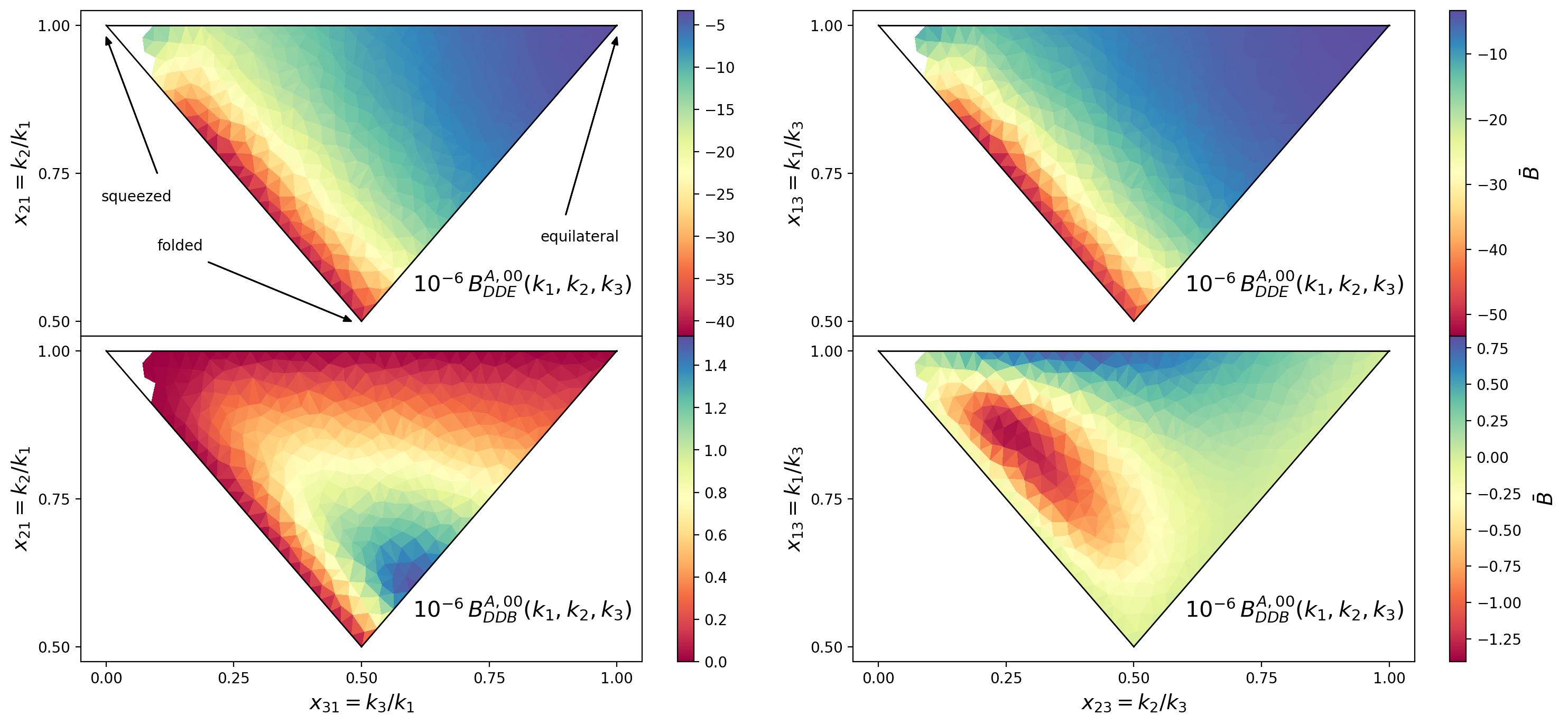}
    \caption{Signal as a function of triangle shape for intrinsic alignment bispectrum multipoles with one single tensor mode. We show two distinct orderings, namely $k_1 \geq k_2 \geq k_3$ (left panel) and $k_3 \geq k_1 \geq k_2$ (right panel). The top (bottom) row shows the parity-even (odd) bispectrum. }
    \label{fig:sst_tri}
\end{figure*}
\begin{figure}[t]
    \centering
    \includegraphics[width=\linewidth]{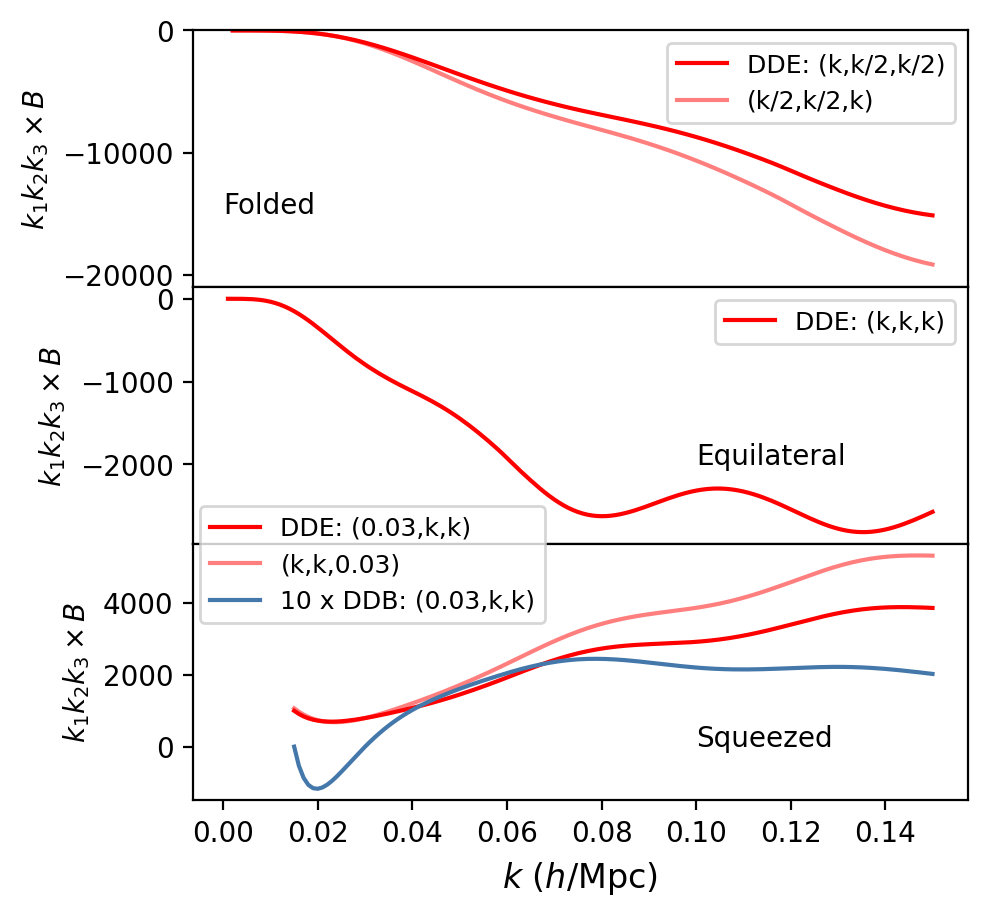}
    \caption{Signal as a function of triangle scale for the first nonzero intrinsic alignment bispectrum multipoles with one tensor mode, i.e. $(\ell_1,\ell_2) = (0,0)$. We show folded, equilateral and squeezed configurations.}
    \label{fig:sst_curv}
\end{figure}
Based on this, we define the ALMs of the bispectrum of shape fields with respect to the associated Legendre polynomials in this paper. Specifically, when one fixes $\hat{\bb{d}} = \hat{\bb{n}}$ then Eq. \eqref{eq:yamamoto} reduces to Eq. \eqref{eq:almdef}. Thus, given that the ALMs are perhaps a more realistic proxy for what a (separable) LPP bispectrum estimator could measure when applied to a galaxy survey, we are motivated to examine the difference between OLMs and ALMs \citep[see][for related work]{inoue}. However, we caution that we still treat the simulation box in the flat-sky approximation (i.e. the distant-observer or GPP limit) whereas a more realistic comparison would involve creating lightcones, applying a survey mask and explicitly incorporating wide-angle geometry. This is beyond the scope of our work.
\subsection{Intrinsic Alignment Bispectra Visualized}
Since we are only interested in the SNR of tensor bispectra, we will omit $B_{DDD}$ and only focus on deterministic parts $B^\text{det,tree}$ in the remainder of this paper. By the results of Section \ref{sec:proj}, we can ignore spectra with more than one B-mode. We thus no longer consider $B_{DDD},B_{DBB}, B_{EBB}$ and $B_{BBB}$ and six combinations out of the 10 in Eq. \eqref{eq:combs} remain. To gain some intuition for the IA bispectrum signal, we present plots for the lowest order nonzero multipoles for all six combinations. This is $(\ell_1,\ell_2) = (0,0),(0,2),(2,2)$ for bispectra with one, two or three tensor fields. We can either use a triangle plot as in \cite{egge_bisp} to visualize the dependence on triangle shape, or fix the geometry of a triangle and plot the signal as a function of scale. For the former, we restrict ourselves to triangle configurations that satisfy $k_l \geq k_m \geq k_n$ and average the bispectrum over the largest triangle side, while keeping the ratios $x_{ml} = k_m/k_l$ and $x_{nl} = k_n/k_l$ fixed. For instance, assuming an ordering $k_1 \geq k_2 \geq k_3$, we have
\begin{equation}
    \bar{B}_{XYZ}^{T,\ell_1,\ell_2}(x_{21},x_{31}) \equiv \int_{k_{\rm min}}^{k_{\rm max}} \mathrm{d}k \, \frac{B_{XYZ}^{T,\ell_1,\ell_2}(k,\,x_{21}\,k,\,x_{31}\,k)}{k_{\rm max} - k_{\rm min}}\,,
\end{equation}
and we average over $k_1$ values between $k_{\rm min} = 0.01\,h\,\mathrm{Mpc}^{-1}$ and $k_{\rm max} = 0.15\,h\,\mathrm{Mpc}^{-1}$. For bispectra involving multiple different fields, one cannot assume without loss of generality that the wavenumbers are ordered as such.  In the following we therefore show different cases, focusing on $k_1 \geq k_2 \geq k_3$ and $k_1 \geq k_3 \geq k_2$ or $k_3 \geq k_1 \geq k_2$, and only the ALMs for the sake of brevity. The fiducial values of bias parameters used to generate these plots are identical to those for the `DESI LRG - like' forecasting setup from Section \ref{sec:forecast}.

When plotting the signal as a function of scale for a fixed triangle shape, we show equilateral, squeezed and collinear configurations, unless they evaluate to zero. For example, parity-odd bispectra evaluate to zero for collinear configurations, as reflections are equivalent to rotations for collinear triangles.

\begin{figure*}[t]
    \centering
    \includegraphics[width=\linewidth]{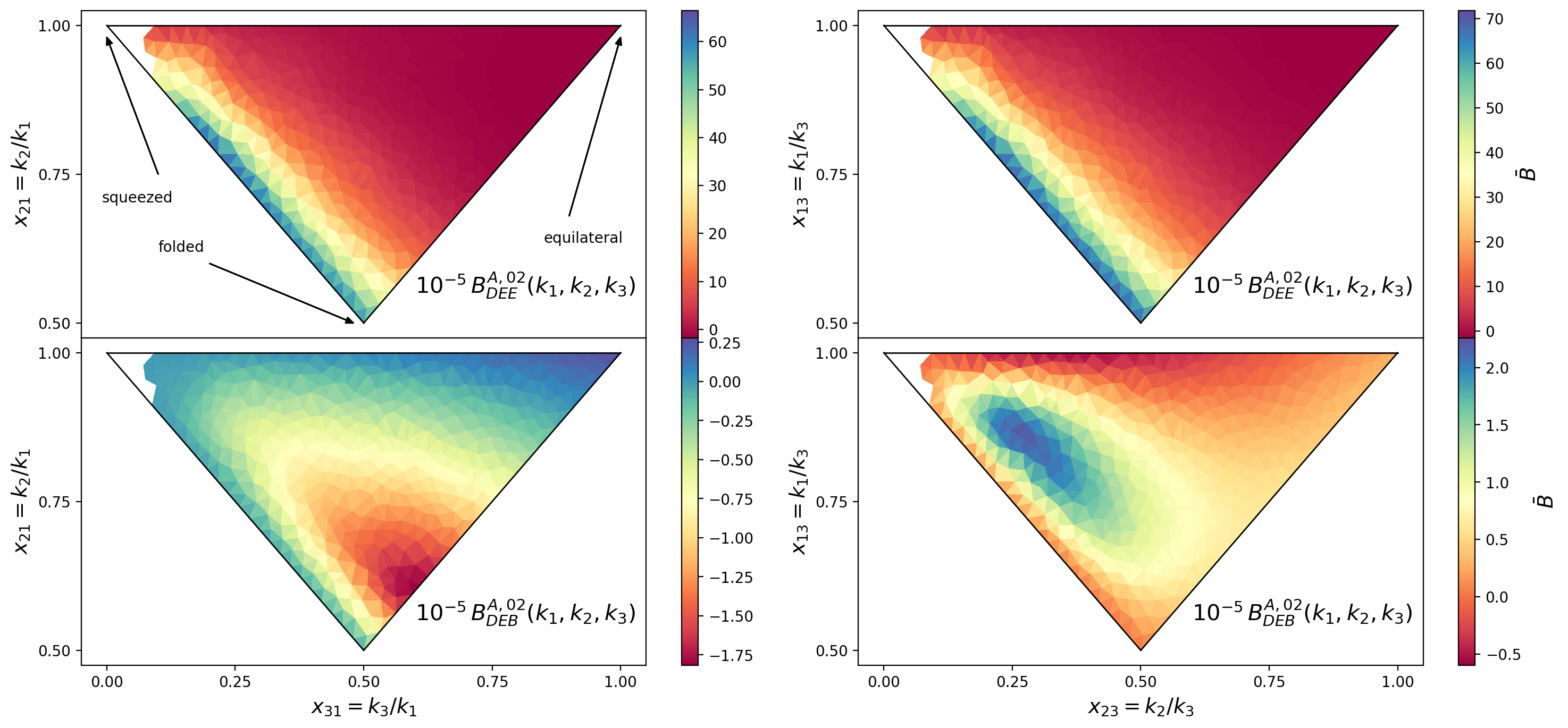}
    \caption{Signal as a function of triangle shape for intrinsic alignment bispectrum multipoles with two tensor modes. We show two distinct orderings, namely $k_1 \geq k_2 \geq k_3$ (left panel) and $k_3 \geq k_1 \geq k_2$ (right panel). The top (bottom) row shows the parity-even (odd) bispectrum.}
    \label{fig:stt_tri}
\end{figure*}

\begin{figure}[t]
    \centering
    \includegraphics[width=\linewidth]{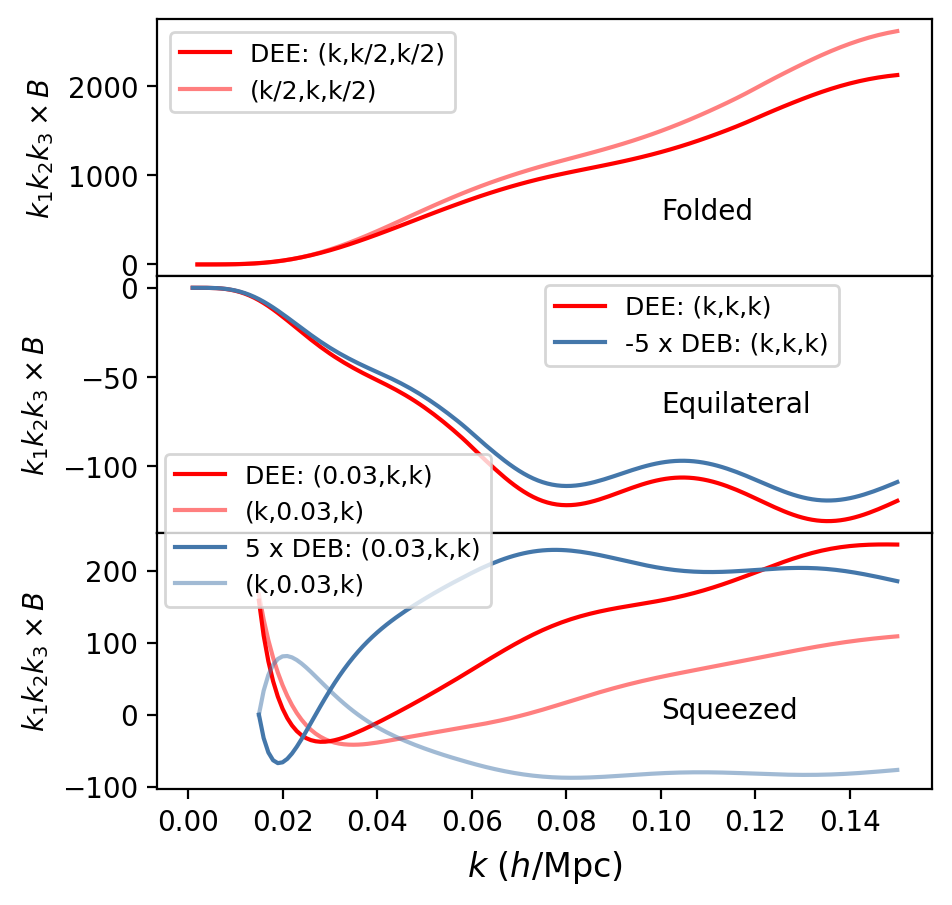}
    \caption{Signal as a function of triangle scale for the first nonzero intrinsic alignment bispectrum multipoles with two tensor modes, i.e. $(\ell_1,\ell_2) = (0,2)$. We show folded, equilateral and squeezed configurations.}
   \label{fig:stt_curv}
\end{figure}

\begin{figure*}[t]
    \centering
    \includegraphics[width=\linewidth]{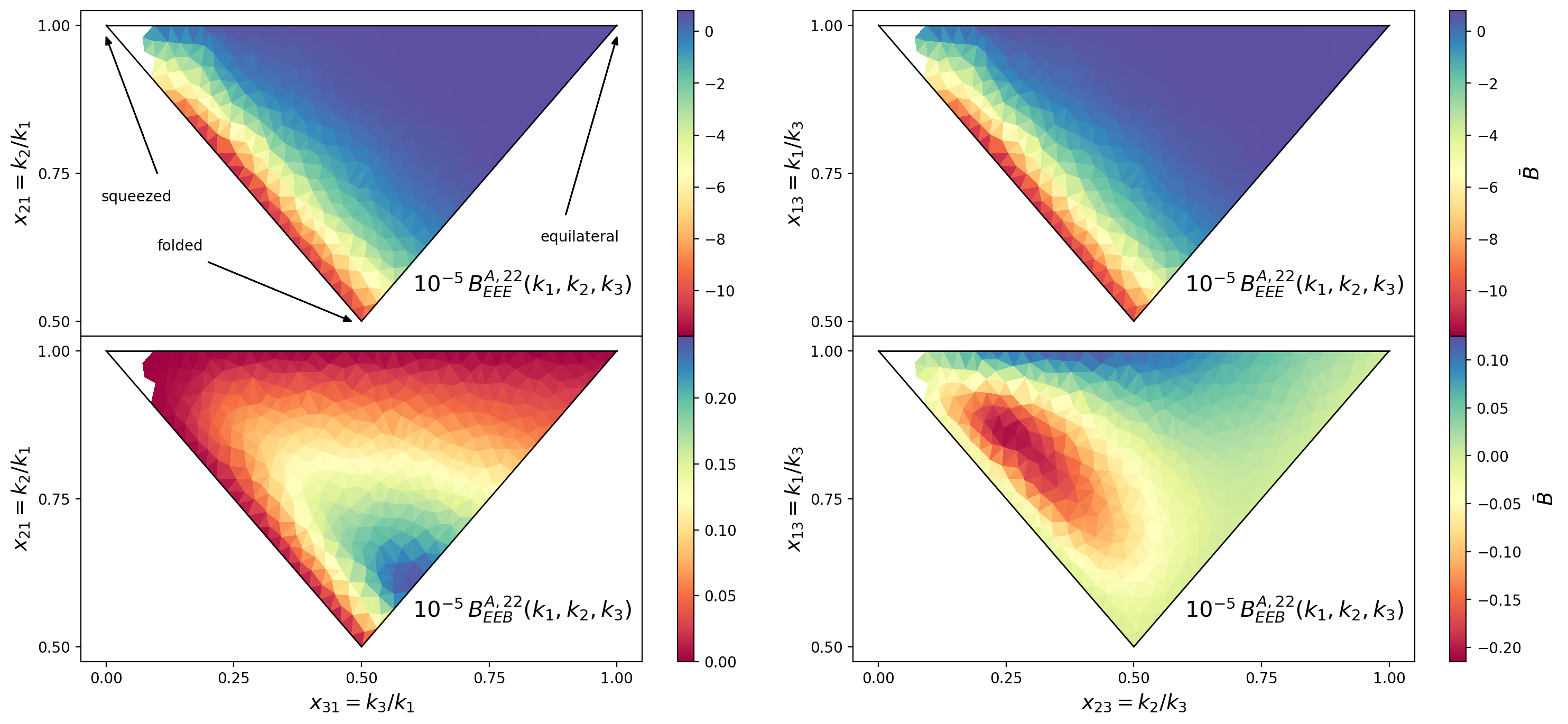}
    \caption{Signal as a function of triangle shape for intrinsic alignment bispectrum multipoles with three tensor modes. We show two distinct orderings, namely $k_1 \geq k_2 \geq k_3$ (left panel) and $k_3 \geq k_1 \geq k_2$ (right panel). The top (bottom) row shows the parity-even (odd) bispectrum.}
    \label{fig:ttt_tri}
\end{figure*}

\begin{figure}[t]
    \centering
    \includegraphics[width=\linewidth]{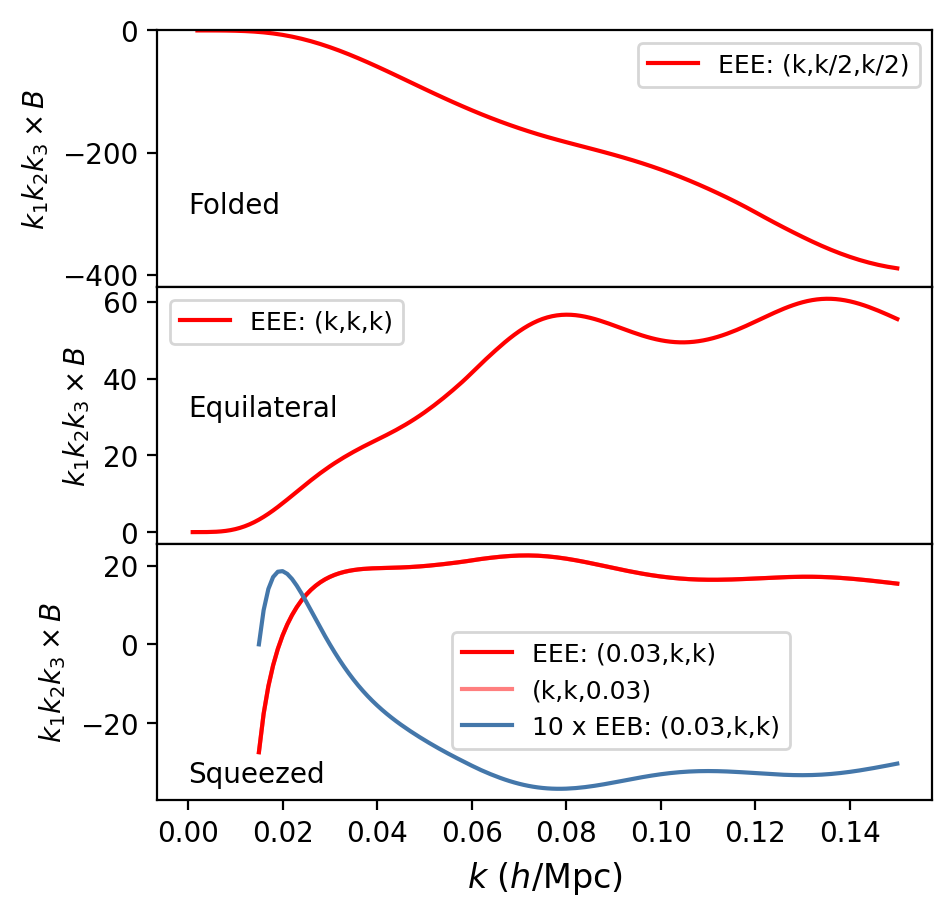}
    \caption{Signal as a function of triangle scale for the first nonzero intrinsic alignment bispectrum multipoles with three tensor modes, i.e. $(\ell_1,\ell_2) = (2,2)$. We show folded, equilateral and squeezed configurations.}
    \label{fig:ttt_curv}
\end{figure}

With this in mind, Figures \ref{fig:sst_tri},\ref{fig:sst_curv} display the signal of the two single-tensor bispectra $B^{00}_{DDE}$ and $B^{00}_{DDB}$. We observe that $B^{00}_{DDB}$ is relatively suppressed compared to $B^{00}_{DDE}$, even though they are na\"ively of the same order in perturbation theory. This was also observed in \cite{schmitz_bisp}; note that in their notation the $s_{ij}$ and $t_{ij}$ terms in the bias expansion dominate the $B_{DDE}$ signal, but these contributions only project onto E-modes\footnote{It may be possible to explain the structure of these triangle plots using analytical arguments \cite{jeong_bisp, smith_bisp}, but we did not pursue this any further.}. Figures \ref{fig:stt_tri} and \ref{fig:stt_curv} show the bispectra with two tensor fields and finally Figures \ref{fig:ttt_tri} and \ref{fig:ttt_curv}, those with three tensor fields.
Note that for $B_{DDB}^{00}$ and $B_{EEB}^{22}$ the signal is antisymmetric in its first two arguments, since swapping the first two sides of the triangle can be achieved by a reflection. Thus, the equilateral configurations evaluate to zero, whereas this is not the case for $B_{DEB}^{02}$. In all cases that involve only the density and $E$-mode fields, the shape dependence of the bispectrum is very similar to that of the auto density bispectrum $B_{DDD}$ \citep[see e.g.,][]{egge_bisp} with an increase in amplitude from equilateral towards collinear triangle configurations (lower left side of the plotted triangular region). For bispectra that also contain a $B$-mode field, the shape dependence is more complex, displaying maxima (or minima) for triangle configurations different from all of the limiting cases. As mentioned above, this is because the bispectrum vanishes for folded and equilateral configurations. As we will see in Section \ref{sec:results}, all spectra with more than one tensor are difficult to detect. As was the case for single-tensor bispectra, replacing an E-mode with a B-mode results in a suppression of the signal. 
\section{Forecasting Setup}\label{sec:forecast}
We are interested in the detectability of the anisotropic large-scale bispectrum of intrinsic alignments. For simplicity, we thus only focus on the SNR of the bispectrum multipoles and defer a joint analysis involving the power spectrum or sensitivity to any specific parameters to future work. The SNR is schematically given by 
\begin{equation}
    (S/N)^2 = \sum_{L,L',\{k_i\}} B^L(\{k_i\}) C_{LL',ij}^{-1} B^{L'}(\{k_j\}) 
\end{equation}
where $L=\ell_1\ell_2$ and $L'=\ell_1'\ell_2'$ are generic multipole indices, $\{k_i\}$ denotes a triangle bin, $C_{LL',ij}$ is the covariance and we suppress the fields $XYZ,X'Y'Z'$ to avoid clutter.
We consider wavenumbers $k_i$ from 15 linearly spaced bins of width $\Delta k = 0.01\,h$/Mpc\footnote{The choice of bin width is not relevant for the SNR; an alternative would be to formally integrate over the space of all triangles as done in e.g. \cite{schmitz_bisp} for their Fisher matrix.} and we choose the lower edge of the first bin to be $k_\text{min} = \Delta k/2$, so that bin centers are integer multiples of $\Delta k$. Our setup thus includes collinear triangle bins, but not open triangle bins (i.e. where the bin centers do not form a closed triangle). The most challenging part of the forecast consists of computing the covariance matrix and its inverse, which we discuss next.
\subsection{Covariance}
We now turn to the calculation of the analytical Gaussian covariance of our multipole estimators. Since our analysis is restricted to large scales, this is a reasonable approach. We will show explicitly that the Gaussian approximation works well on the scales we consider here \cite{paper2}, which is in line with findings of \cite{oddo_bisp}. Another potential avenue for cross-checking the validity of the analytic Gaussian covariance would be via injecting intrinsic alignment into N-body simulations \citep[e.g., as in][]{vanalfen}.

As per the approach in \cite{bakx_eft}, we will proceed to calculate the Gaussian covariance by plugging in the NLA model \cite{bridle_king} predictions (including shape noise $c^\text{g}$) for the power spectra (see Appendix \ref{sec:cov}). They are repeated below for convenience \footnote{We use here the non-linear power spectrum purely to remain consistent with the choice in \cite{bakx_eft}. However, on the scales we consider here the difference between the nonlinear and linear power spectrum does not exceed 10 percent and can thus be considered negligible.}: 
\begin{equation}\label{eq:nlamod}
\begin{aligned}
    P_{EE}(k,\mu) &= \frac{1}{4}(b_1^{\text{g}})^2(1-\mu^2)^2P_{\text{NL}}(k) + c^{\text{g}}; \\
    P_{DE}(k,\mu) &= \frac{1}{2}b_1^\text{s}b_1^{\text{g}}(1-\mu^2)P_{\text{NL}}(k); \\
    P_{BB}(k,\mu) &= c^{\text{g}}.
\end{aligned}
\end{equation}
For the galaxy power spectrum we also use a linear bias model: 
\begin{equation}
    P_{DD}(k) = (b_1^\text{s})^2 P_\text{NL}(k) + c^\text{s}.
\end{equation}
Here, $b_1^\text{s}$ is the familiar linear galaxy bias parameter and we take $c^\text{s} = 1/\bar{n}$ where $\bar{n}$ is the number density of objects in the sample. Furthermore, $P_{\text{NL}}(k)$ is the nonlinear matter power spectrum, which we calculate using the \texttt{CCL} wrapper for \texttt{HaloFit} \cite{ccl,halofit}. To a good approximation, $c^{\text{g}}$ is just the typical ellipticity of the galaxy sample, i.e. $c^{\text{g}} = \sigma_\gamma^2 / \bar{n}$ \cite{bakx_eft,kurita_iapower}. The linear alignment parameter $b_1^{\text{g}}$ can be expressed in terms of the more commonly used $A_\text{IA}$ via \cite{bakx_eft}
\begin{equation}
    b_1^{\text{g}} = -2A_\text{IA} C_1 \rho_\text{cr} \Omega_{m,0} / D(z),
\end{equation}
where $D(z)$ is the linear growth factor and $C_1\rho_\text{cr} = 0.0134$ \cite{hirata_seljak,bridle_king}. It is understood that $b_1^{\text{g}}$ and $A_\text{IA}$ may also be redshift dependent. It was shown in related work \cite{hymalaia} that this approximation works fairly well on large scales for the power spectrum covariance. If desired, the bispectrum-bispectrum and power spectrum-trispectrum terms could be included in the bispectrum covariance by using perturbation theory \cite{floss_cov,biagetti_cov,sefusatti_bisp2}.

In the following, we always keep the ordering of the tracers (i.e. $XYZ$) fixed. This does not lose any generality, cf. Eq. \eqref{eq:ordering}. Because the tracers under consideration need not be the same ($X,Y,Z$ can be any of $D, E,B$), the multipoles $B_{XYZ}^{T,\ell_1\ell_2}(k_1,k_2,k_3)$ are in general \textit{not} symmetric with respect to their arguments, even when $\ell_1 = \ell_2 = 0$. This needs to be taken into account when computing the covariance matrix associated with the bispectrum data vector. We cannot immediately impose an ordering on the triangle bins such as $k_1\geq k_2 \geq k_3$, as one would typically do for galaxy clustering -- this would in principle result in a loss of information (unless one then also considers all permutations of fields $XYZ$, which yields equivalence to our approach).

The result of the covariance calculation for an arbitrary cross-correlation of different multipoles, different tracers and different triangle bins can be found in Eq. \eqref{eq:covresult}. In particular, we can infer from this expression that two bispectra evaluated in isomorphic triangles (e.g. $(k_1,k_2,k_3)$ and $(k_2,k_3,k_1)$) are not identical, but they have a nontrivial cross correlation. This makes the inversion of even the Gaussian covariance matrix more difficult than for the clustering bispectrum (we elaborate on this below). To circumvent this issue, we can also consider \textit{symmetrized observables} $\widetilde{B}$ which sum over all isomorphic triangle configurations. Explicitly,
\begin{equation}\label{eq:altdef}
    \widetilde{B}_{XYZ}^{T,\ell_1\ell_2}(k_1,k_2,k_3) =\frac{1}{6}\sum_{\sigma \in S_3} B^{T,\ell_1\ell_2}_{XYZ}(k_{\sigma(1)},k_{\sigma(2)},k_{\sigma(3)}).
\end{equation}
For the symmetrized observables, we can by construction assume that $k_1 \geq k_2 \geq k_3$.  Note that depending on $X,Y$ and $Z$ some terms in this average may be identical. The covariance of $\widetilde{B}$ is diagonal in $k$-space. However, the symmetrized estimate is essentially a form of data compression (it cannot be inverted) and thus in general lowers the SNR, as mentioned above. Quantifying the difference in SNR between $\widetilde{B}$ and $B$ is one of the objectives of this Section. 

As a leading example, consider the covariance matrix of $B^{00}_{DDE}(k_1,k_2,k_3).$ This multipole is symmetric in its first two arguments and so we must impose $k_1 \geq k_2$ to avoid overcounting. It consists of three parts, namely $C^{\text{gen}}$ for general triangles (i.e. with unequal sides), $C^{\text{isos}}$ for isosceles triangles and $C^{\text{equi}}$ for equilateral ones. Each of these has to be inverted separately. Indeed these types of triangles are uncorrelated, and the full covariance takes the following block diagonal form: 
\begin{equation}\label{eq:totcov}
    C = \begin{pmatrix}
        C^{\text{gen}} & 0 & 0 & 0 \\
        0 & C^{\text{isos},1} & 0 & 0 \\
        0 & 0 & C^{\text{isos},2} & 0 \\
        0 & 0 & 0 & C^{\text{equi}} 
    \end{pmatrix}.
\end{equation}
We consider $C^{\text{gen}}$ first. 
Denoting wavenumber bins with integers for simplicity, we have that e.g. bin $(4,3,2)$ is covariant with bin $(4,2,3)$ and bin $(3,2,4)$, but we need not consider the other three permutations, because they do not satisfy $k_1>k_2$. Accordingly, $C^{\text{gen}}$ takes a 3 by 3 block form. For isosceles triangles, there are two distinct cases: in the first block, the equal sides are the longest (i.e. $(3,3,2), (3,2,3)$ since both should still satisfy $k_1 \geq k_2$), or in the second block the equal sides are the shortest, for example $(2,2,3), (3,2,2)$. This makes $C^{\text{isos},i}$ a $2 \times 2$ block matrix for $i=1,2$. We thus have
\begin{equation}\label{eq:blockcov}
    C^{\text{gen}} = \begin{pmatrix}
        C_{11} & C_{12} &C_{13} \\
        C_{12}^T & C_{22} &C_{23} \\
        C_{13}^T & C_{23}^T & C_{33}
    \end{pmatrix}; \quad 
    C^{\text{isos},i} = \begin{pmatrix}
        C_{11}^i & C_{12}^i \\
        C^{i,T}_{12} & C_{22}^i 
    \end{pmatrix};
\end{equation}
where each of the $C_{ab},C_{ab}^i$ is diagonal. 
Finally, $C^{\text{equi}}$ is already diagonal and its inversion is trivial. 

If we consider more multipoles (be it different $\ell_1,\ell_2$ or different $X,Y,Z$) which are also symmetric in their first two arguments, then the only difference is that every entry in the above matrix is replaced by an $m \times m$ matrix, where $m$ is the number of multipoles considered. The process of inverting this matrix remains virtually unchanged, because the component matrices $C_{ij}$ from Eq. \eqref{eq:blockcov} are themselves still block diagonal. If the bispectrum has no exchange symmetries, as is the case for e.g. $B_{D E B}^{00}$ or $B_{DDE}^{20}$, analytically inverting the covariance would still be possible but we do not consider this case here. Finally, note that in the Gaussian approximation, there is no cross-correlation between parity-even and parity-odd bispectra. This makes it particularly easy to combine e.g. $B_{DDE}$ and $B_{DDB}$ if desired.

\subsection{Analysis Choices}
We need to estimate the values of the EFT parameters that enter the bispectrum predictions. For this, in our baseline analysis we will employ so-called Lagrangian co-evolution relations \cite{bakx_eft,akitsu_bias}, which assume that initially, a linear bias relation holds for the tracer in question and non-linear bias arises only through non-linear advection, such that we only need to specify the values of the linear alignment bias $b_1^{\text{g}}$ and the linear galaxy bias $b_1^{\text{s}}$. In modelling the signal, we set all of the stochastic parameters to zero so that $B^\text{stoch,tree}=0$\footnote{We will see in future work that the stochastic bispectrum is in fact nonzero. However, it does not contribute any information beyond what is already present in the power spectrum and for that reason we neglect it here.}. 
We neglect redshift uncertainties throughout, i.e. we require a spectroscopic sample in order to resolve the three-dimensional positions of the galaxies needed to construct the multipoles. We do not incorporate redshift-space distortions, but note that our multipole expansions and corresponding estimators are applicable also when they are included. We employ the Gaussian covariance from the previous subsection throughout. 

Our default setup consists of a sample of luminous red galaxies (LRGs) from the Dark Energy Spectroscopic Instrument (DESI) \cite{desisb} where we assume to have imaging from the Legacy Survey of Space and Time (LSST) \cite{lsst}. Given the sky coverage $f_\text{sky}$ and the redshift range of the survey, we can compute the total (comoving) survey volume as
\begin{equation}
    V_\text{survey} = \frac{4}{3}\pi f_\text{sky}(\chi(z_\text{max})^3 - \chi(z_\text{min})^3)
\end{equation}
By default, we use $A_{\text{IA}} = 4$ from \cite{kurita_png} and a volume-limited sample for $0.4 < z < 0.8$ with number density $\bar{n} = 5\times 10^{-4} (h/\text{Mpc})^3$ \cite{zhou}. We evaluate the theory at a mean redshift of $z_*=0.6$. For the sky coverage we assume an area of $\sim 4,000$ square degrees ($f_\text{sky}=0.1$). This yields a total of $\sim 1.3$ million LRGs. The linear galaxy bias is estimated to be roughly $1.5/D(z_*) \approx 2.05$\cite{chaussidon_png}. We use a fiducial maximum wavenumber of $k_{\rm{max}}=0.14\,h$/Mpc. The RMS shape dispersion is $\sigma_\gamma = 0.18$ (per component).

We expect the highest SNR to come from $B_{DDE}^{00}$ (both for the OLMs and the ALMs), which is nonzero even if all the higher-order shape biases vanish (c.f. Table \ref{tab:params}). Therefore, in the case of $B_{DDE}^{00}$ specifically, we also assess the difference between using a symmetrized estimator $\widetilde{B}$ as in Eq. \eqref{eq:altdef} and using the full estimator which does not symmetrize over any of the wavevector arguments.  

The expressions for the OLMs in general need to be integrated numerically; this is due to the presence of factors of $1-\mu_i^2$ in the denominator of Eq. \eqref{eq:phasefac}\footnote{This phase factor fortuitously cancels when computing the theory expression for the power spectrum, so there this issue does not arise.}. By contrast, the expressions for the ALMs and the covariance matrix from Eq. \eqref{eq:covresult} are simple polynomials in $\cos \theta_1,\sin \theta_1$ and $\cos \xi,\sin \xi$; they can be integrated analytically using the identities given in Appendix \ref{sec:angint}.
\section{Results}\label{sec:results}
\begin{figure}
    \centering
    \includegraphics[width=\linewidth]{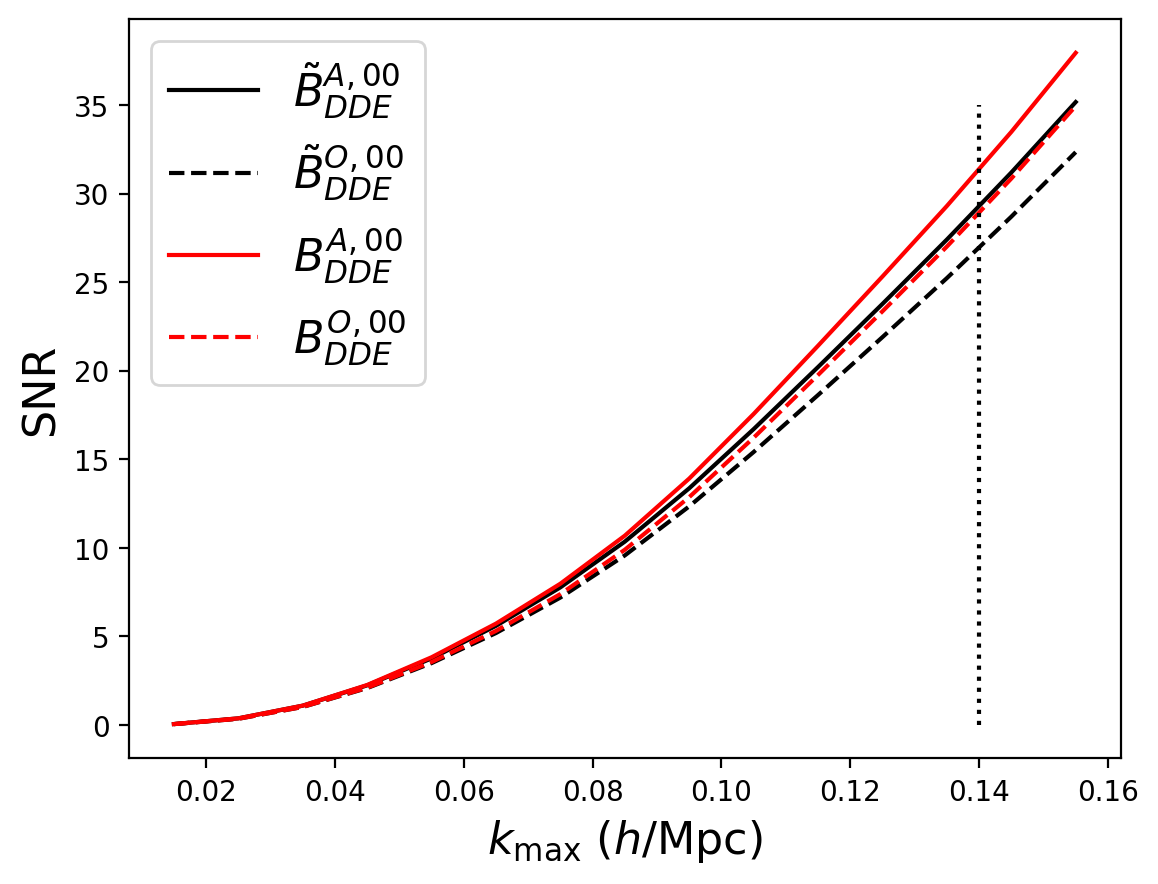}
    \caption{Cumulative signal-to-noise ratio as a function of maximum wavenumber for the lowest order associated and ordinary multipoles of $B_{DDE}$. Solid curves are ALMs, which always exceed the SNR of the OLMs (dashed curves). Red curves are unsymmetrized, which always exceed the SNR in the symmetrized case (black).}
    \label{fig:gge_snr}
\end{figure}

We first assess the detectability of $B_{DDE}^{00}$. The SNR is highest if we do not compress the data vector by symmetrizing. Furthermore, we are interested in the difference between the use of OLMs and ALMs. The SNR for each of these four cases as a function of scale cut is plotted in Figure \ref{fig:gge_snr}. The OLM expansion of the bispectra should strictly speaking contain the same amount of information as the ALM expansion. However, in practice the convergence of either series may differ when considering only a finite number of multipole moments \cite{inoue}. We see in this case that the SNR in the OLMs is consistently lower than that of the ALMs. Moreover, while the SNR in the unsymmetrized data vector is higher, the difference is small ($<10\%$). At $k_\text{max}=0.14\,h/$Mpc, indicated by the dashed vertical line, we obtain a promising SNR of $\approx 25-30$ in all cases. 

\begin{figure}
    \centering
    \includegraphics[width=\linewidth]{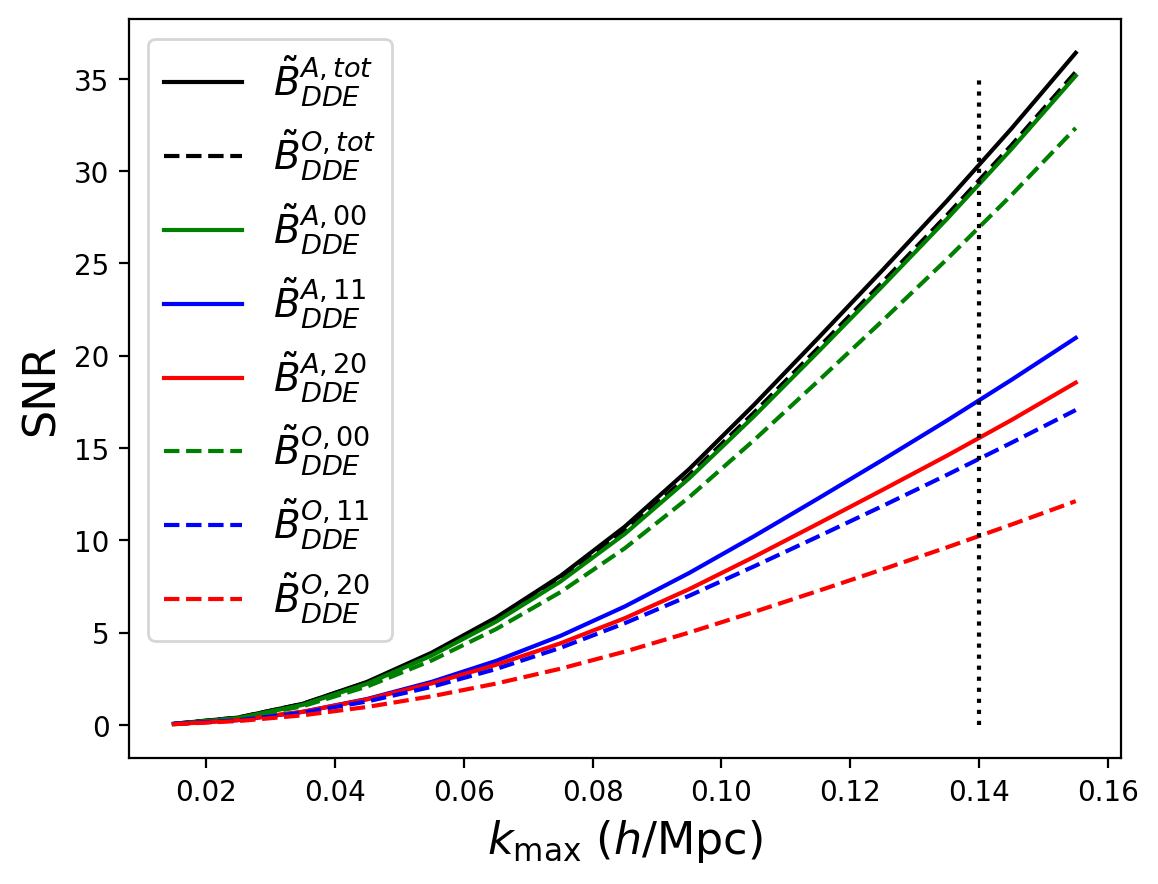}
    \caption{Cumulative signal-to-noise ratio as a function of maximum wavenumber for several symmetrized associated and ordinary multipoles of $B_{DDE}$. Solid (dashed) lines represent ALMs (OLMs) and black lines represent the SNR in the sum of the three multipoles.}
    \label{fig:ggemults_snr}
\end{figure}

Motivated by these results, we also examine the SNR in symmetrized higher-order multipoles of $B_{DDE}$, specifically $(\ell_1,\ell_2) = (1,1)$ and $(\ell_1,\ell_2) = (2,0)$\footnote{Note that also for symmetrized multipoles, one can without loss of generality keep the order of the fields fixed without discarding any further information, e.g. any multipole of $\tilde{B}_{EDD}$ can always be rewritten as a linear combination of multipoles of $\tilde{B}_{DDE}$ by virtue of Eq. \eqref{eq:ordering} and the fact that $\mu_3$ is linearly dependent on $\mu_1,\mu_2$. }. The result is shown in Figure \ref{fig:ggemults_snr}. We predict significant detections for both these multipoles, at SNR $\approx 15$ at the fiducial scale cut for the ALMs. Again, the ALMs outperform the OLMs, with large differences for fixed $(\ell_1,\ell_2)$. However, the difference in the SNR obtained when combining these three multipoles (black lines) is very small. For completeness, we also display in Table \ref{tab:snrvals} the SNR values for several specific triangle configurations. Interestingly, while for $(\ell_1,\ell_2) = (0,0)$ and $(2,0)$ the SNR peaks in the collinear configurations (third and fourth rows), it is the squeezed limit that contains the most SNR for $(\ell_1,\ell_2) = (1,1)$.
\begin{table}
    \centering
    \begin{tabular}{|c|c|c|c|}
    \hline
        Bin $(h/\text{Mpc})$ / $(\ell_1,\ell_2)$ & $(0,0)$ &$ (1,1)$ & $(2,0)$ \\
    \hline 
    \hline
        $(0.05,0.05,0.05)$ & 0.29 & 0.34 & 0.22\\
    \hline 
        $(0.10,0.10,0.10)$ & 0.33 & 0.39 & 0.26\\
    \hline 
        $(0.15,0.10,0.05)$ & 3.22 & 0.51 & 2.43\\
    \hline 
        $(0.14,0.07,0.07)$ & 2.60 & 0.40 & 1.96\\
    \hline 
        $(0.15,0.15,0.10)$ & 0.69 & 0.73 & 0.41\\
    \hline 
        $(0.15,0.15,0.03)$ & 1.07 & 0.92 & 0.37\\
    \hline
    \end{tabular}
    \caption{The SNR in specific triangle bins for different ALMs of $\tilde{B}_{DDE}$.}
    \label{tab:snrvals}
\end{table}

\begin{figure}
    \centering
    \includegraphics[width=\linewidth]{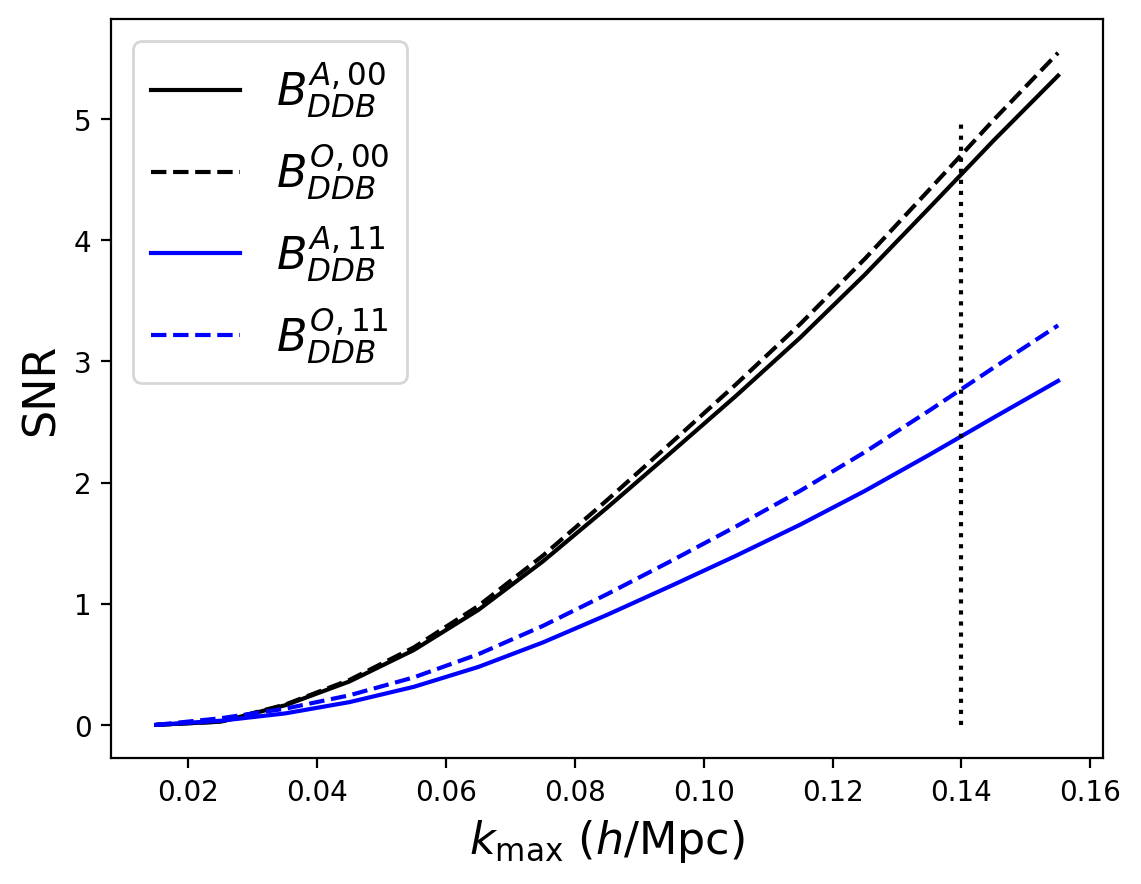}
    \caption{Cumulative signal-to-noise ratio as a function of maximum wavenumber for two associated and ordinary multipoles of $B_{DDB}$, namely $(\ell_1,\ell_2) = (0,0)$ (black) and $(\ell_1,\ell_2) = (1,1)$ (blue).}
    \label{fig:ggb_snr}
\end{figure}

Next, we examine two multipoles of $B_{DDB}$, namely $(\ell_1,\ell_2) = (0,0)$ and $(\ell_1,\ell_2) = (1,1)$ in Figure \ref{fig:ggb_snr}. 
Here we find an important difference to that of $B_{DDE}^{00}$: after imposing $k_1 \geq k_2$, the (Gaussian) covariance is already diagonal in $k$-space, by virtue of the fact that $P_{GB}(\bb{k})=0$ (one can easily check this using Eq. \eqref{eq:covresult}). Thus, we do not symmetrize these multipoles (this symmetrization would also lead to zero signal since these bispectra are antisymmetric in their first two arguments). We see that the SNR of these bispectra is significantly suppressed compared to the parity-even case of $B_{DDE}$, consistent with the findings of Section \ref{sec:proj}. However, in this case the OLM expansion seems to outperform the ALM expansion. Still, the overall detection significance is more modest at SNR $\approx 5 \,(2)$ for $(\ell_1,\ell_2) = (0,0)$ resp. $(\ell_1,\ell_2) = (1,1)$.

\begin{figure}
    \centering
    \includegraphics[width=\linewidth]{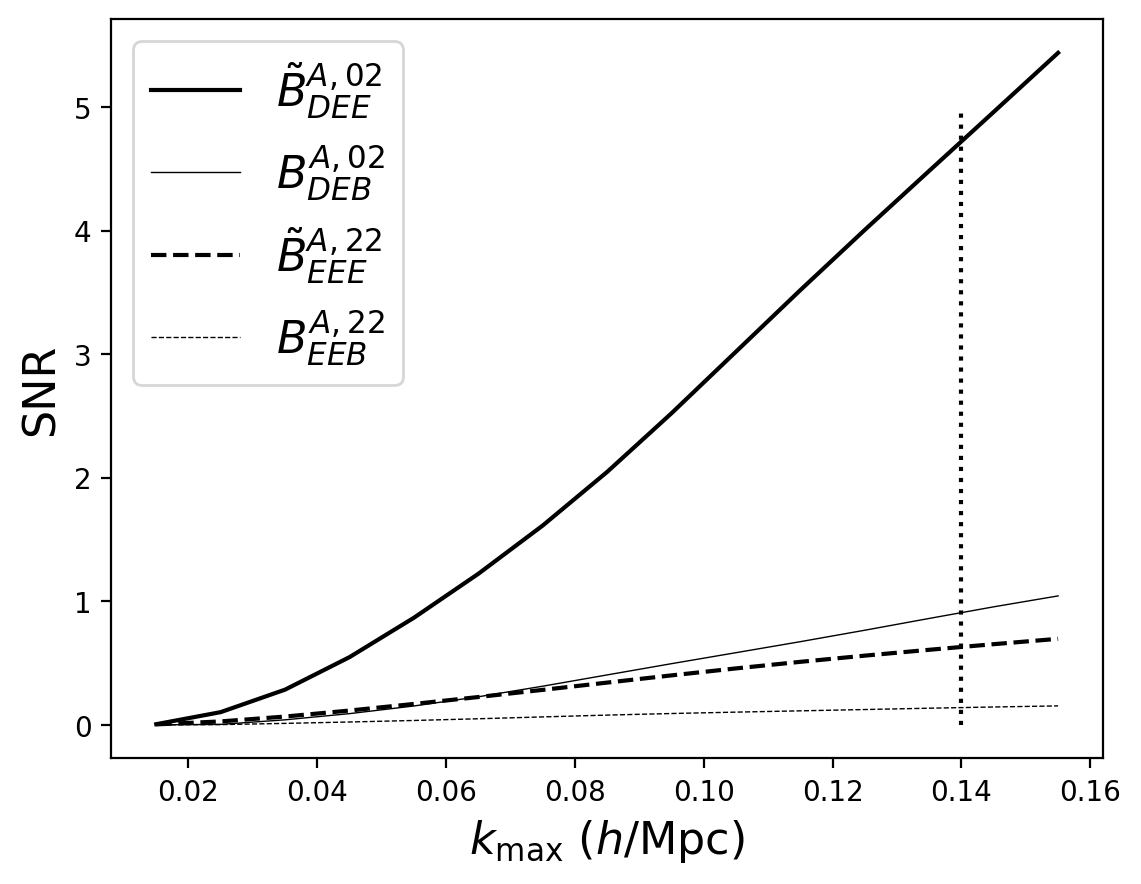}
    \caption{Cumulative signal-to-noise ratio as a function of maximum wavenumber for the lowest order associated multipoles of bispectra with more than one tensor field. Note that the odd bispectra are not symmetrized.}
    \label{fig:tt_snr}
\end{figure}

For the spectra with more than one tensor field, obtaining a detection is more challenging with our fiducial setup. Figure \ref{fig:tt_snr} displays the SNR for the lowest order multipoles of bispectra with two or three shape fields as a function of the scale cut $k_\text{max}$\footnote{We decided not to symmetrize the parity-odd bispectra, since this lead to a significant drop in SNR.}. Only $B_{DEE}$ seems detectable, at moderate significance at best (SNR $\approx 5$). The reason for this difference is similar to the situation for the alignment auto-power spectrum: since the linear alignment bias is very small $(b_1^\text{g}<0.1)$, the signal is noise-dominated essentially at all scales; $(b_1^\text{g})^2 P_L(k) < \sigma_\gamma^2/\bar{n}$. This is even more exacerbated for the pure-tensor bispectra $B_{EEE}$ and $B_{EEB}$. Obtaining a sufficiently dense and aligned galaxy sample so as to detect these may be possible with samples that possess stronger intrinsic alignment or higher number density (see \cite{shi_ia} for a potential way to increase the IA amplitude). In addition, it may be possible to detect these spectra in N-body simulations, in which case they provide additional consistency checks. 

\section{Discussion}\label{sec:disc}
We have shown in this work that (i) arbitrary bispectra of tensor fields can be modelled consistently within the framework of the effective field theory of intrinsic alignments; (ii) the resulting projected bispectra are either parity-even or parity-odd depending on the number of B-modes they contain; (iii) their estimation from spectroscopic galaxy samples via multipole decomposition and separable FFT-based algorithms is analogous to the existing literature on multipoles of the galaxy bispectrum in redshift space, both in the parity-even and the parity-odd case, and finally (iv) single-tensor parity-even bispectra are detectable on large scales with high SNR in Stage IV galaxy surveys, some of which are already operational. 

Our results can be extended or improved upon in various ways. Notably, it may be possible to increase the SNR by considering different `tensor tracers' than LRG intrinsic alignment, such as the recently introduced `multiplet alignment' from \cite{lamman_multiplet} or by employing different estimators that increase the IA amplitude \cite{shi_ia}. Second, the sky overlap with LSST is still not completely certain; in our fiducial setup the SNR is simply proportional to $\sqrt{f_\text{sky}}$ so that increasing it by a factor of $2$ would yield $\sim 40\%$ more SNR. Moreover, if one does not restrict to imaging from LSST but rather uses DESI Legacy imaging, shapes would be available for the entire DESI footprint (see \cite{kurita_png} for a similar approach; the imaging quality requirements for intrinsic alignment measurements may not be as stringent as those for shear measurements). Moreover, there may also be significant overlap with the 4MOST galaxy survey in the southern hemisphere \cite{4most}. These enhancements may also yield a positive detection of bispectra with more tensor fields. Crucially, we have also set all stochastic contributions to bispectra to zero and only considered the deterministic part of the bias expansion when computing the SNR. It may however still be possible to extract information from these contributions; we leave this for future work. It is also possible to extend the analysis presented here to smaller scales, which would appear to increase the SNR significantly. While the magnitude of the signal may not differ that much from the tree-level prediction beyond $k=0.14\,h$/Mpc, the covariance receives non-Gaussian contributions from higher order terms \cite{biagetti_cov} and super-sample effects \cite{kurita_iapower,takada_hu} and is therefore considerably more uncertain. For the power spectrum of intrinsic alignments, \cite{kurita_iapower} showed that this causes the cumulative SNR to eventually plateau. 

Our findings could be extended to other three-point observables such as the three-point correlation function (3PCF) \cite{schneider_lombardi1,schneider_lombardi2} or aperture mass statistics \cite{kilbinger_ap}. It would be valuable to quantify the increase in constraining power from retaining modes outside the plane of the sky ($\mu\neq 0$), for example by using wedges as done for the galaxy power spectrum \cite{chen_lagr}. Moreover, statistics that are sensitive to specific triangle configurations may not extract all possible information from the alignment signal. For example, the equal-radius aperture mass statistic used in \cite{linke_lowz} is mostly sensitive to equilateral triangles \cite{kilbinger_ap}, while the SNR for the bispectrum multipoles (see Table \ref{tab:snrvals}) does not peak there.

We anticipate that accurate knowledge of large-scale intrinsic alignment can help mitigate potential systematic selection effects that contaminate growth rate measurements via redshift-space distortions \cite{hirata_gr,krause_bisp}. Moreover, the bispectrum is a useful probe in constraining primordial non-Gaussianity \cite{cabass_mf,cabass_sf}. Consistently including the IA bispectrum in analyses constraining PNG would involve extending our modelling to include redshift-space distortions \cite{taruya_rsd,okumura_rsd}. Furthermore, it would be interesting to explore the sensitivity of the IA bispectrum to any of the phenomena examined in the context of the IA power spectrum, examples of which include magnetic fields \cite{akitsu_magn}, modified gravity \cite{harvey} or evolving dark energy \cite{okumura_gr}. In addition, parity-odd bispectra (i.e. $B_{TTB}$) have proven useful in constraining graviton parity and Gaussianity using the Cosmic Microwave Background (CMB) temperature and polarization data \cite{philcox_grav} - it is in principle possible to also obtain bounds on primordial physics from LSS tracers \cite{cabass_collider1,cabass_collider2}. Generally, IA can be used to place generic bounds on tensor perturbations via their impact on large-scale tidal fields \cite{philcox_shapes,schmidt_shear,gorji}. 

With regards to weak gravitational lensing, the perturbative nature of our specific model may seem to restrict its applicability on small scales. However, \cite{chen_lensing} shows that perturbative models for IA \cite{chen_ia} can still be used to extract the amplitude of matter fluctuations $S_8 = \sigma_8 \sqrt{\Omega_m/0.3}$ provided that one makes appropriate scale cuts. It would be very interesting to include three-point statistics in such an analysis. Moreover, the tree-level bispectrum model employed here correctly captures the angular and scale dependence of the anisotropic tensor bispectrum at least at leading order \cite{paper2}; it may be possible to formulate phenomenological extensions of it (in a similar vein to \cite{linke_lowz,pyne_illustris}) that extend to smaller scales, in the same way that the NLA model extends the linear alignment model from \cite{catelan_la} to smaller scales \cite{bridle_king}.

We have restricted ourselves to a calculation of only the SNR, rather than performing a full forecast for any specific parameter. In order to properly assess the sensitivity of the tensor bispectrum to any given parameter of interest in the Fisher matrix formalism \cite{tegmark_fisher}, one requires knowledge of the corresponding derivative with respect to that parameter. It may be that the inclusion of the tensor bispectrum helps to break degeneracies between parameters, in which case the information gain exceeds what is expected from simply computing the SNR (relative to that of other statistics). We leave these considerations for future work. Lastly, in an upcoming follow-up paper \cite{paper2} we will explicitly verify our theoretical modelling against N-body simulations, as done in \cite{bakx_eft}.
\begin{acknowledgements}
    We thank Fabian Schmidt for useful discussions throughout this project. This publication is part of the project ``A rising tide: Galaxy intrinsic alignments as a new probe of cosmology and galaxy evolution'' (with project number VI.Vidi.203.011) of the Talent programme Vidi which is (partly) financed by the Dutch Research Council (NWO). AE is supported at the Argelander Institut f\"ur Astronomie by an Argelander Fellowship. Z.V. acknowledges the support of the Kavli Foundation.
\end{acknowledgements}
\newpage
\appendix
\section{Explicit EFT expressions}\label{sec:eft}
Here we collect formulae for the tree-level tensor bispectrum in the EFT of IA. We first discuss the deterministic contributions and then stochastic terms, before summarizing the results. All expressions can be straightforwardly generalized to cases where not all shape tracers are the same, but we do not do this here to avoid clutter. 
\subsection{Deterministic Bispectrum Contributions}
Here we provide explicit expressions for the deterministic part of the bispectrum $B^{\text{det,tree}}$. At tree-level, it contains only terms that involve four factors of the linear density contrast $\delta^{(1)}$.  When the symmetric 2-tensor $S_{ij}$ is expanded in perturbation theory, it takes the form 
\begin{equation}
    S_{ij}(\mathbf{x}) = \sum_{n=0}^\infty S_{ij}^{(n)}(\mathbf{x})
\end{equation}
where the $n$-th order term in Fourier space is 
\begin{equation}
    S_{ij}^{(n)}(\mathbf{k}) = \int \md^3\bb{p}_1 \dots \md^3\bb{p}_n\, \mathcal{K}_{ij}^{(n)}(\mathbf{p}_1, \dots \mathbf{p}_n)\, \delta^D(\mathbf{k}-\mathbf{p}_{1n}) \delta^{(1)}(\mathbf{p}_1) \dots \delta^{(1)}(\mathbf{p}_n).
\end{equation}
where $\bb{p}_{1n} = \bb{p}_1 + \dots + \bb{p}_n$. Here the bias-dependent PT kernel $\mathcal{K}_{ij}^{(n)}$ splits into a trace and trace-free part: 
\begin{equation}
    \mathcal{K}_{ij}^{(n)} = \mathcal{K}_{ij}^{\text{s},(n)}+\mathcal{K}_{ij}^{\text{g},(n)}.
\end{equation}
The total bispectrum of shape tracers at tree-level is then calculated to be 
\begin{equation}\label{bispec}
\begin{aligned}
B_{ijklrs}^{\alpha \beta \gamma, \text{det,tree}}(\mathbf{k}_1,\mathbf{k}_2,\mathbf{k}_3) 
&= 2\,\mathcal{K}_{ij}^{\alpha,(1)}(\mathbf{k}_1)\mathcal{K}_{kl}^{\beta,(1)}(\mathbf{k}_2)\mathcal{K}_{rs}^{\gamma,(2)}(\mathbf{k}_1,\mathbf{k}_2)P_L(k_1)P_L(k_2)+ \text{2 perm.}; 
\end{aligned}    
\end{equation}
where $\alpha,\beta,\gamma$ are either s or g and the overall factor of $2$ is a consequence of Wick's theorem. The first order kernels are given by
\begin{equation}
\begin{aligned}
    \mathcal{K}_{ij}^{\text{s},(1)}(\mathbf{k}) &= \frac{1}{3}\delta_{ij}b_1^{\text{s}}; \\
    \mathcal{K}_{ij}^{\text{g},(1)}(\mathbf{k}) &= b_1^{\text{g}} \left( \frac{\bb{k}_i\bb{k}_j}{k^2}-\frac{1}{3}\delta_{ij}\right) = b_1^{\text{g}}\mathcal{D}_{ij}(\bb{k});
\end{aligned}
\end{equation}
where in the second line we introduced the trace-free Fourier space operator $\mathcal{D}_{ij}(\bb{k}) = \bb{k}_i\bb{k}_j/k^2-\frac{1}{3}\delta_{ij} $.
Expressions for the second order kernels are
\begin{equation}\label{eq:secker}
    \begin{aligned}
        \mathcal{K}_{ij}^{\text{s},(2)}(\mathbf{k}_1,\bb{k}_2) &= \frac{1}{3}\delta_{ij}\bigg[ b_1^{\text{s}} F_2(\mathbf{k}_1,\bb{k}_2) + b_{2,1}^{\text{s}}\bigg(\frac{(\bb{k}_1\cdot \bb{k}_2)^2}{k_1^2k_2^2} \bigg) + b_{2,2}^{\text{s}} \bigg];\\
        \mathcal{K}_{ij}^{\text{g},(2)}(\mathbf{k}_1,\bb{k}_2) &= \text{TF}\bigg[c_{1}^\text{g} \bigg(\frac{\bb{k}_{12,i}\bb{k}_{12,j}}{k_{12}^2}F_2(\mathbf{k}_1,\bb{k}_2)\bigg) + c_{2,1}^{\text{g}}\bigg(\frac{\bb{k}_{12,i}\bb{k}_{12,j}}{k_{12}^2}F_2(\mathbf{k}_1,\bb{k}_2)-\frac{1}{2}\frac{\bb{k}_1\cdot \bb{k}_2}{k_1^2k_2^2}(\bb{k}_{1,i}\bb{k}_{1,j}+\bb{k}_{2,i}\bb{k}_{2,j})\bigg) \\
        &+ c_{2,2}^{\text{g}}\bigg(\frac{1}{2}\frac{\bb{k}_1\cdot \bb{k}_2}{k_1^2k_2^2}(\bb{k}_{1,i}\bb{k}_{2,j}+\bb{k}_{2,i}\bb{k}_{1,j})\bigg) + c_{2,3}^{\text{g}} \bigg( \frac{\bb{k}_{1,i}\bb{k}_{1,j}}{2k_1^2}+\frac{\bb{k}_{2,i}\bb{k}_{2,j}}{2k_2^2}\bigg)\bigg];
        \end{aligned}
\end{equation}
where $\bb{k}_{12} = \bb{k}_1+\bb{k}_2$ and the $F_2$ kernel equals the standard SPT expression 
\begin{equation}
    F_2(\mathbf{k}_1,\bb{k}_2) = \frac{5}{7}+\frac{1}{2}\frac{\bb{k}_1\cdot \bb{k}_2}{k_1 k_2}\bigg( \frac{k_1}{k_2}+\frac{k_2}{k_1}\bigg)+\frac{2}{7}\frac{(\bb{k}_1\cdot \bb{k}_2)^2}{k_1^2k_2^2}.
\end{equation}
Here and in our companion paper as well as \cite{vlah_eft1,bakx_eft}, we choose to work with a redefinition of bias coefficients which is given by 
\begin{equation}
\begin{aligned}
    b_1^{\text{s}} &= c_1^\text{s}; \qquad b_{2,1}^{\text{s}} = c_{2,2}^{\text{s}} + \frac{2}{7} c_{2,1}^{\text{s}}; \qquad b_{2,2}^{\text{s}} = c_{2,3}^{\text{s}} + \frac{5}{7}c_{2,1}^{\text{s}}; \\  
    b_1^{\text{g}} &=c_1^\text{g}; \qquad b_{2,1}^{\text{g}} = c_{2,1}^{\text{g}}+c_{2,2}^{\text{g}}+c_{2,3}^{\text{g}}; \qquad b_{2,2}^{\text{g}} = c_{2,3}^{\text{g}}+\frac{20}{7}c_{2,1}^{\text{g}}; 
    \qquad b_{2,3}^{\text{g}}=c_{2,3}^{\text{g}}.    
\end{aligned}
\end{equation}
The coefficients $c_{2,a}^\text{s}$ can be related to other familiar bases by using the relations in App. B of \cite{schmidt_bias}. In the notation of \cite{mirbabayi}, the coefficients $c_1^{\{\text{s,g}\}},c_{2,a}^{\{\text{s,g}\}} (a=1,2,3)$ belong to trace and trace-free parts of the operators
\begin{equation}
    \Pi^{[1]}_{ij}, \Pi^{[2]}_{ij},(\Pi^{[1]})^2_{ij}, \Pi^{[1]}_{ij}\text{Tr}(\Pi^{[1]}_{ij})
\end{equation} 
respectively - thus, there are only two independent bias coefficients for the trace part. By contrast, the trace-free part contains three free parameters. Importantly, it is insufficient to write an expansion for the total symmetric tensor field $S_{ij}$ and then equate the coefficients occurring in the trace and trace-free parts \cite{vlah_eft1}. This can ultimately be explained as a consequence of renormalization of the bare coefficients.

We will need the co-evolution relations for the second order parameters in terms of $b_1^{\text{s}}$ and $b_1^{\text{g}}$, which are derived under the assumption that the Lagrangian higher order bias parameters are zero. They read \cite{bakx_eft}
\begin{equation}
    b_{2,1}^{\text{g}} = (b_1^{\text{s}}-1)b_1^{\text{g}}; \qquad b_{2,2}^{\text{g}} = (b_1^{\text{s}}-20/7)b_1^{\text{g}}; \qquad b_{2,3}^{\text{g}} = b_1^{\text{s}}b_1^{\text{g}}.
\end{equation}
We will also need the co-evolution relations for the scalar bias parameters, which read \cite{schmidt_bias}
\begin{equation}
    b_{2,1}^{\text{s}} = b_{K^2} = -\frac{2}{7}(b_1^{\text{s}}-1); \qquad b_{2,2}^{\text{s}} = b_{\delta^2}-\frac{1}{3}b_{K^2} = \frac{10}{21}(b_1^{\text{s}}-1).
\end{equation}
This part of the bispectrum already appeared in \cite{schmitz_bisp}, albeit in a less general setting. Our treatment also differs from theirs in terms of the stochastic contributions, which were not considered in their work. We describe these next.
\subsection{Stochastic Contributions}
We now discuss the stochastic contributions to the scalar and shape kernels. For the scalar part we have two terms contributing to the \textit{galaxy} bias expansion (both of which are absent for dark matter). They are well known and read 
\begin{equation}
    \delta_D^\text{stoch}(\bb{x}) \qquad \supset \qquad \epsilon(\mathbf{x}) \text{ (first order) }; \qquad \epsilon^\delta(\bb{x})\delta(\bb{x}) \text{ (second order) }.
\end{equation}
Moving on to the shape part, there are two new independent bias operators contributing to the tree-level bispectrum. They are 
\begin{equation}
    g_{ij}^{\text{stoch}}(\bb{x}) \qquad \supset \qquad \epsilon_{ij}(\mathbf{x}) \text{ (first order) }; \qquad \epsilon_{ij}^\delta(\bb{x})\delta(\bb{x}),\quad \epsilon^\Pi(\mathbf{x})\text{TF}(\Pi^{[1]})_{ij}(\bb{x}) \text{ (second order) }.
\end{equation}
The relevant stochastic contractions we can form with these fields (that is, using at most one second order operator and omitting combinations that come with only  one instance of  $\delta(\bb{x})$) are \footnote{Strictly speaking, the scale by which the corrections to stochastic spectra are suppressed is in general different from $k_\text{NL}$, but we will ignore this difference here (since these corrections are ignored altogether in this work).}
\begin{equation}
    \begin{aligned}
        \langle \epsilon(\bb{k}_1)\epsilon(\bb{k}_2)\rangle' &= \alpha_{\text{ss}}^{11} + \mathcal{O}((k/k_{\text{NL}})^2);\\
        \langle \epsilon(\bb{k}_1)\epsilon^\delta(\bb{k}_2)\rangle' &= \alpha_{\text{ss}}^{1\delta} + \mathcal{O}((k/k_{\text{NL}})^2);\\
        \langle \epsilon(\bb{k}_1)\epsilon_{ij}^\delta(\bb{k}_2)\rangle' &= 0 + \mathcal{O}((k/k_{\text{NL}})^2);\\
        \langle \epsilon_{ij}(\bb{k}_1)\epsilon(\bb{k}_2)\rangle' &= 0 + \mathcal{O}((k/k_{\text{NL}})^2);\\
        \langle \epsilon_{ij}(\bb{k}_1)\epsilon^\Pi(\bb{k}_2)\rangle' &= 0 + \mathcal{O}((k/k_{\text{NL}})^2);\\
        \langle \epsilon_{ij}(\bb{k}_1)\epsilon^\delta(\bb{k}_2)\rangle' &= 0 + \mathcal{O}((k/k_{\text{NL}})^2);\\
        \langle \epsilon(\bb{k}_1)\epsilon^\Pi(\bb{k}_2)\rangle' &= \alpha_{\text{sg}}^{1\Pi} + \mathcal{O}((k/k_{\text{NL}})^2);\\
        \langle \epsilon_{ij}(\mathbf{k}_1)\epsilon_{kl}(\mathbf{k}_2)\rangle ' &= \alpha_{\text{gg}}^{11}\Delta_{ij,kl}^{\text{TF}}  + \mathcal{O}((k/k_{\text{NL}})^2);\\
        \langle \epsilon_{ij}(\mathbf{k}_1)\epsilon^\delta_{kl}(\mathbf{k}_2)\rangle ' &= \alpha_{\text{gg}}^{1\delta}\Delta_{ij,kl}^{\text{TF}}  + \mathcal{O}((k/k_{\text{NL}})^2);\\\langle\epsilon(\bb{k}_1)\epsilon(\bb{k}_2)\epsilon(\bb{k}_3)\rangle' &= \alpha_\text{sss}^{111} + \mathcal{O}((k/k_{\text{NL}})^2); \\
        \langle \epsilon(\bb{k}_1)\epsilon(\bb{k}_2)\epsilon_{ij}(\bb{k}_3)\rangle' &= 0 + \mathcal{O}((k/k_{\text{NL}})^2); \\
        \langle \epsilon(\bb{k}_1)\epsilon_{ij}(\bb{k}_2)\epsilon_{kl}(\bb{k}_3)\rangle' &= \alpha_{\text{sgg}}^{111}\Delta_{ij,kl}^{\text{TF}}  + \mathcal{O}((k/k_{\text{NL}})^2); \\
        \langle \epsilon_{ij}(\mathbf{k}_1)\epsilon_{kl}(\mathbf{k}_2)\epsilon_{rs}(\mathbf{k}_3)\rangle ' &=\alpha_{\text{ggg}}^{111}\Delta_{ij,kl,rs}^{\text{TF}} + \mathcal{O}((k/k_{\text{NL}})^2).    \end{aligned}
\end{equation}
Here the $\alpha$ parameters are stochastic amplitudes, which we label according to the nature of the fields being correlated (scalar s or tensor g) and the specific stochastic operator label $1,\Pi,\delta$ for $\epsilon,\epsilon^\delta,\epsilon^\Pi$ respectively. There are five in total for the tensor sector (i.e. where at least one tensor field is involved) and three for the pure scalar sector. The stochastic amplitude $\alpha_\text{gg}^{11}$ also occurred in \cite{bakx_eft} and is related to the shape-noise amplitude in the EE- and BB- power spectra $c^\text{g}$ from Eq. \eqref{eq:nlamod} through $2\alpha_\text{gg}^{11}=c^\text{g}$.
The symbol $\Delta^{\text{TF}}$ is the total pair-symmetric, pair-traceless tensor: it is symmetric under exchange of pairs, and traceless symmetric in each pair. It is a uniquely defined expression in terms of the Kronecker delta (up to an irrelevant constant). 
The total the pair-symmetric, pair-traceless tensor with four indices is given by 
\begin{equation}
\Delta^{\text{TF}}_{ij,kl} = \delta_{ik}\delta_{jl}+\delta_{il}\delta_{jk}-\frac{2}{3}\delta_{ij}\delta_{kl}.
\end{equation}
The total pair-symmetric, pair-traceless tensor with six indices is given by \cite{vlah_eft1}
\begin{equation}
    \begin{aligned}
        \Delta_{ij,kl,rs}^{\text{TF}}&=\frac{4}{3}\delta_{ij}\delta_{kl}\delta_{rs}+ \frac{3}{4}\bigg(\delta_{ik}\delta_{jr}\delta_{ls}+\delta_{ik}\delta_{js}\delta_{lr}+\delta_{il}\delta_{jr}\delta_{ks}+\delta_{il}\delta_{js}\delta_{lr}+\delta_{ir}\delta_{jk}\delta_{ls}+\delta_{ir}\delta_{js}\delta_{kl}\\
        &+\delta_{is}\delta_{jk}\delta_{lr}+\delta_{is}\delta_{jr}\delta_{kl}\bigg) - \bigg(\delta_{ij}(\delta_{kr}\delta_{ls}+\delta_{ks}\delta_{lr})+\delta_{kl}(\delta_{ir}\delta_{js}+\delta_{is}\delta_{jr})+\delta_{rs}(\delta_{il}\delta_{jk}+\delta_{ik}\delta_{jl})\bigg).\\
    \end{aligned}
\end{equation}
For scalar spectra we recover the familiar combinations
\begin{equation}
    \begin{aligned}
        B^\text{sss} \supset \langle \epsilon(\bb{k}_1)\epsilon(\bb{k}_2)\epsilon(\bb{k}_3)\rangle' &= \alpha_\text{sss}^{111} + \mathcal{O}((k/k_{\text{NL}})^2); \\
        B^\text{sss} \supset \langle (b_1^\text{s}\delta)(\bb{k}_1)\epsilon(\bb{k}_2)(\epsilon^\delta \delta)(\bb{k}_3)\rangle' &= b_1^\text{s}\alpha_\text{ss}^{1\delta}P_L(k_1) + \mathcal{O}((k/k_{\text{NL}})^2); \\
    \end{aligned}
\end{equation}
where the second line should be symmetrized over $\bb{k}_1,\bb{k}_2$ and $\bb{k}_3$. The relevant spectra containing tensor fields take the form\footnote{Note that the contribution on the second line was missing in \cite{vlah_eft1}.} 
\begin{equation}\label{eq:3dstoch}
\begin{aligned}
    B_{ij}^{\text{ssg}} &\supset \langle \epsilon(\bb{k}_1)(b_1^\text{s}\delta)(\bb{k}_2)(\epsilon^\Pi \text{TF}(\Pi^{[1]})_{kl})(\bb{k}_3)\rangle' = b_1^\text{s} \alpha_\text{sg}^{1\Pi} \mathcal{D}_{kl}(\bb{k}_2)P_L(k_2) + \mathcal{O}((k_i/k_{\text{NL}})^2);\\
    B_{ij}^{\text{ssg}} &\supset \langle \epsilon(\bb{k}_1)(\epsilon^\delta \delta)(\bb{k}_2)(b_1^\text{g}\text{TF}(\Pi^{[1]})_{kl})(\bb{k}_3)\rangle' = b_1^\text{g} \alpha_\text{ss}^{1\delta} \mathcal{D}_{kl}(\bb{k}_3)P_L(k_3) + \mathcal{O}((k_i/k_{\text{NL}})^2);\\
    B_{ijkl}^{\text{sgg}} &\supset \langle \epsilon(\bb{k}_1)(b_1^\text{g}\text{TF}(\Pi^{[1]})_{ij})(\bb{k}_2)(\epsilon^\Pi \text{TF}(\Pi^{[1]})_{kl})(\bb{k}_3)\rangle' = b_1^\text{g}\alpha_\text{sg}^{1\Pi} \mathcal{D}_{ij}(\bb{k}_2) \mathcal{D}_{kl}(\bb{k}_3)P_L(k_2) + \mathcal{O}((k_i/k_{\text{NL}})^2);\\
    B_{ijkl}^{\text{sgg}} &\supset \langle \epsilon(\bb{k}_1)\epsilon_{ij}(\mathbf{k}_2)\epsilon_{kl}(\mathbf{k}_3)\rangle ' = \alpha_{\text{sgg}}^{111} \Delta_{ij,kl}^{\text{TF}} + \mathcal{O}((k_i/k_{\text{NL}})^2);\\
    B_{ijkl}^{\text{sgg}} &\supset \langle (b_1^\text{s}\delta)(\bb{k}_1)\epsilon_{ij}(\mathbf{k}_2)(\epsilon_{kl}^\delta\delta)(\mathbf{k}_3)\rangle ' = b_1^\text{s} \alpha_{\text{gg}}^{1\delta} \Delta_{ij,kl}^{\text{TF}}P_L(k_1) + \mathcal{O}((k_i/k_{\text{NL}})^2);\\
    B_{ijklrs}^{\text{ggg}} &\supset \langle (b_1^\text{g}\text{TF}(\Pi^{[1]})_{ij})(\bb{k}_1)(\epsilon_{kl}^\delta\delta)(\mathbf{k}_2)\epsilon_{rs}(\mathbf{k}_3)\rangle ' = b_1^\text{g}\alpha_{\text{gg}}^{1\delta} \mathcal{D}_{ij}(\bb{k}_1) \Delta_{kl,rs}^{\text{TF}}P_L(k_1)+ \mathcal{O}((k_i/k_{\text{NL}})^2); \\
    B_{ijklrs}^{\text{ggg}} &\supset \langle \epsilon_{ij}(\mathbf{k}_1)\epsilon_{kl}(\mathbf{k}_2)\epsilon_{rs}(\mathbf{k}_3)\rangle ' = \alpha_{\text{ggg}}^{111} \Delta_{ij,kl,rs}^{\text{TF}}+ \mathcal{O}((k_i/k_{\text{NL}})^2).
\end{aligned}
\end{equation}
Care must be taken to symmetrize the above expressions over the indices and wavevectors. In particular, for the first line, there is also a contribution to $B^{\text{ssg}}$ where $\bb{k}_2$ is replaced with $\bb{k}_1$. Furthermore, lines 2 and 5 should be multiplied by a factor of 2 to obtain the total contribution and on line 3 there is a contribution with $\bb{k}_2$ and $\bb{k}_3$ exchanged. The sixth line should be summed over all six permutations of $\bb{k}_1, \bb{k}_2$ and $\bb{k}_3$. One may also observe that $\alpha_\text{ss}^{11}$ and $\alpha_\text{gg}^{11}$ do not contribute to tree-level bispectra. Instead, they only occur in two-point statistics at leading order. 

The $\mathcal{O}((k_i/k_{\text{NL}})^2)$ terms indicate scale-dependent corrections to the constant amplitudes $\alpha$ as well as $k$-dependent tensorial structures such as
\begin{equation}    \alpha_\text{ggg}^{\Pi}\left( k_{\text{NL}}^{-2}(\bb{k}_{1,i}\bb{k}_{1,j}-\frac{1}{3}\delta_{ij}k_1^2)
    \Delta_{kl,rs} + 2 \text{ perm.}\right); 
\end{equation}
(for the last line of Eq. \eqref{eq:3dstoch} specifically) and many more, which necessarily start at higher order since $\hat{\bb{k}}_i$ is not allowed by locality in space.

Thus, the total tensorial bispectrum (before projecting shapes into E- and B-fields) contains a total of 13 free parameters at tree-level (excluding higher-derivative corrections), namely 
\begin{equation}\label{eq:allparams}
    \underbrace{b_1^{\{\text{s,g}\}}, b_{2,1}^{\{\text{s,g}\}}, b_{2,2}^{\{\text{s,g}\}}, b_{2,3}^{\text{g}}}_{\text{bias expansion (7)}},\quad \underbrace{\alpha_\text{ss}^{1\delta},\alpha_\text{sg}^{1\Pi},\alpha_\text{gg}^{1\delta},\alpha_\text{sss}^{111},\alpha_\text{sgg}^{111},\alpha_\text{ggg}^{111}}_{\text{stochasticity (6)}}.
\end{equation} It actually turns out that we do not need to consider $\alpha_\text{ggg}^{111}$ in the analysis of projected shapes, as we show below. 
It is easy to verify that 
\begin{equation}\label{eq:stochshape}
\begin{aligned}
\bb{M}^{X}_{ij}(\hat{\bb{k}}_1) \bb{M}^{Y}_{kl}(\hat{\bb{k}}_2)\Delta^\text{TF}_{ij,kl} &= 2 \text{ Tr}(\bb{M}^{X}(\hat{\bb{k}}_1) \bb{M}^{Y}(\hat{\bb{k}}_2)); \\
    \bb{M}^{X}_{ij}(\hat{\bb{k}}_1) \bb{M}^{Y}_{kl}(\hat{\bb{k}}_2)\bb{M}^{Z}_{rs}(\hat{\bb{k}}_3)\Delta^\text{TF}_{ij,kl,rs} &= 3 \text{ Tr}(\bb{M}^{X}(\hat{\bb{k}}_1) \bb{M}^{Y}(\hat{\bb{k}}_2)\bb{M}^{Z}(\hat{\bb{k}}_3)) + 3 \text{ Tr}(\bb{M}^{Y}(\hat{\bb{k}}_2) \bb{M}^{X}(\hat{\bb{k}}_1)\bb{M}^{Z}(\hat{\bb{k}}_3)). 
\end{aligned}
\end{equation}
By the cyclicity of the trace, these two expressions are fully symmetric in $X,Y,Z$. The expression on the first line can be evaluated for the different combinations of E- and B-fields as 
\begin{equation}
\begin{aligned}
    2 \text{ Tr}(\bb{M}^{E}(\hat{\bb{k}}_2) \bb{M}^{E}(\hat{\bb{k}}_3)) &= \cos 2(\phi_{2} - \phi_{3});  \\
    2 \text{ Tr}(\bb{M}^{E}(\hat{\bb{k}}_2) \bb{M}^{B}(\hat{\bb{k}}_3)) &= \sin 2(\phi_{2} - \phi_{3}); \\
    2 \text{ Tr}(\bb{M}^{B}(\hat{\bb{k}}_2) \bb{M}^{B}(\hat{\bb{k}}_3)) &= \cos 2(\phi_{2} - \phi_{3}). \\
\end{aligned}   
\end{equation}
As expected, they only depend on the differences between azimuthal angles $\phi_i$. The expressions on the right-hand side can be expressed in terms of $\mu_1$ and $\xi$. Specifically, we have
\begin{equation}\label{eq:sincosphi}
\begin{aligned}
    \cos 2\phi_{2} &= \frac{\mu_{12}^2(1-\mu_1^2) + 2\mu_1\mu_{12}\sqrt{(1-\mu_1^2)(1-\mu_{12}^2)}\cos \xi - (1-\mu_{12}^2)(\sin^2\xi - \mu_1^2\cos^2\xi) }{1-\mu_2^2}; \\
    \sin 2\phi_{2} &=\frac{2\sqrt{1-\mu_{12}^2}\big(\mu_{12}\sqrt{1-\mu_1^2}+\mu_1\sqrt{1-\mu_{12}^2}\cos \xi\big)\sin \xi}{1-\mu_2^2}; \\
    \cos 2\phi_{3} &= \frac{(1-\mu_1^2)(k_1+k_2\mu_{12})^2+ k_2\big(2\mu_1(k_1+2k_2\mu_{12})\sqrt{(1-\mu_1^2)(1-\mu_{12}^2)}\cos \xi -k_2(1-\mu_{12}^2)(\sin^2\xi - \mu_1^2\cos^2\xi)\big)}{k_3^2(1-\mu_3^2)};\\
    \sin 2\phi_{3} &= \frac{2k_2\big((k_1+k_2\mu_{12})\sqrt{(1-\mu_1^2)(1-\mu_{12}^2)}+k_2\mu_1(1-\mu_{12}^2)\cos\xi \big)\sin \xi}{k_3^2(1-\mu_3^2)};
\end{aligned}
\end{equation}
where 
\begin{equation}\label{eq:mu2mu3rel}
    \begin{aligned}
        \mu_2 &= \mu_1 \mu_{12} - \sqrt{(1-\mu_1^2)(1-\mu_{12}^2)}\cos \xi; \\
        \mu_3 &= -\frac{\mu_1(k_1+k_2\mu_{12})+k_2\sqrt{(1-\mu_1^2)(1-\mu_{12}^2)}\cos \xi}{k_3};
    \end{aligned}
\end{equation}
all assuming without loss of generality that $\phi_{1}=0$. Remarkably, the expression in the second line of Eq. \eqref{eq:stochshape} always vanishes for every combination of E- and B-fields. Hence, the three-point function of the stochastic shape noise $\epsilon_{ij}(\bb{x})$ does not contribute to spectra of projected quantities. Nevertheless, this contribution, corresponding to $\alpha_\text{ggg}^{111}$, is present at the 3D level. 
\subsection{Total Bispectrum}

Here we summarize broadly the structure of the total expressions for the tree-level bispectrum. As mentioned in Section \ref{sec:proj}, all contributions to spectra with a single B-mode are odd in $\xi$, that is, $B_{DDB},B_{DEB}$ and $B_{EEB}$. This is because changing $\xi$ to $-\xi$ amounts to reflecting the triangle in the $\bb{k}_1-\hat{\bb{n}}$ plane, and these spectra are odd under such reflections (which, again, does \textit{not} mean that these spectra are zero, because the triangle does not get mapped onto itself). Moreover, since 
\begin{equation}   \bb{M}_{ij}^B(\hat{\bb{k}})\mathcal{K}_{ij}^{\text{g},(1)}(\bb{k}) = b_1^{\text{g}} \,\bb{M}_{ij}^B(\hat{\bb{k}})\mathcal{D}_{ij}(\bb{k})\delta(\bb{k}) = 0,
\end{equation}
the B-mode starts at second order, and all contributions to $B_{BBB}$ vanish at tree level. We hence do not include it anymore. There are however stochastic contributions to $B_{DBB}$ and $B_{EBB}$. It is instructive to tabulate all the dependencies of the bispectra on the list of parameters that occur at tree level. Here we use the letter $D$ to indicate a generic \textit{scalar} tracer, e.g. galaxy number counts. Note that $\alpha_{\text{ss}}^{11},\alpha_{\text{gg}}^{11}$ and $\alpha_{\text{ggg}}^{111}$ have been omitted; none of the bispectra depend on these parameters. Thus, the multipoles $B_{XYZ}$ with $X,Y,Z$ = $D,E,B$ depend on 12 free parameters in total (seven deterministic, five stochastic). The dependencies are indicated in Table \ref{tab:params} below.
\begin{table}
\begin{center}
\begin{tabular}{|c|c|c|c|c|c|c|c|c|c|c|c|c|c}
\hline
    $B$ / Bias & $b_1^{\text{s}}$ & $b_{2,1}^{\text{s}}$ & $b_{2,2}^{\text{s}}$ & $b_1^{\text{g}}$ & $b_{2,1}^{\text{g}}$ & $b_{2,2}^{\text{g}}$ & $b_{2,3}^{\text{g}}$ &  $\alpha_\text{ss}^{1\delta}$ & $\alpha_\text{sg}^{1\Pi}$ & $\alpha_\text{gg}^{1\delta}$ & $\alpha_\text{sss}^{111}$ &$\alpha_\text{sgg}^{111}$\\
    \hline
    $B_{G G G}$ & \cm & \cm & \cm & $\crs$ & $\crs$ & $\crs$ & $\crs$ & \cm & $\crs$ & $\crs$ & \cm & $\crs$ \\
    \hline
    $B_{G G E}$ & \cm & \cm & \cm & \cm & \cm & \cm & \cm & \cm & \cm & $\crs$ & $\crs$ & $\crs$ \\
    \hline
    $B_{G G B}$ & \cm & $\crs$ & $\crs$ & $\crs$ & \cm &  $\crs$ & \cm & $\crs$ & \cm & $\crs$ & $\crs$ & $\crs$ \\
    \hline
    $B_{G E E}$ & \cm & \cm & \cm & \cm & \cm & \cm & \cm & $\crs$ & \cm & \cm  &  $\crs$  & \cm  \\
    \hline
    $B_{G B B}$ & $\crs$ & $\crs$ & $\crs$ & $\crs$ & $\crs$ & $\crs$ & $\crs$ & $\crs$  & $\crs$ & \cm & $\crs$ & \cm   \\
    \hline
    $B_{G E B}$ & \cm & $\crs$ & $\crs$ & \cm & \cm &  $\crs$ & \cm &  $\crs$ & $\crs$ & \cm & $\crs$ & \cm  \\
    \hline
    $B_{E E E}$ & $\crs$ & $\crs$ & $\crs$ & \cm & \cm & \cm & \cm &  $\crs$ & $\crs$ & \cm  & $\crs$ & $\crs$   \\
    \hline
    $B_{E E B}$ & $\crs$ & $\crs$ & $\crs$ & \cm & \cm &  $\crs$ & \cm  & $\crs$ & $\crs$ & \cm & $\crs$ & $\crs$ \\
    \hline
    $B_{E B B}$ & $\crs$ & $\crs$ & $\crs$ & $\crs$ & $\crs$ & $\crs$ & $\crs$ & $\crs$ & $\crs$ &\cm & $\crs$ & $\crs$ \\
    \hline
\end{tabular}
\caption{Parameter dependence of tensorial bispectra.\label{tab:params}}
\end{center}
\end{table}
There can be additional `accidental' degeneracies when only considering specific multipoles at tree level; for example in the case of $B_{DDB}$ one can easily check that all OLMs and ALMs of odd multipoles depend only on $ b_{2,1}^{\text{g}}$ and $b_{2,3}^{\text{g}}$, and furthermore the lowest order ALM $(\ell_1=\ell_2=0,m_Z=2)$, namely 
\begin{equation}
    B_{DDB}^{A,00}(k_1,k_2,k_3) \propto \frac{1}{4\pi}\int \md \mu_1 \md \xi\, (1-\mu_3^2) \bb{M}_{ij}^B(\hat{\bb{k}}_3)\mathcal{K}_{ij}^{\text{g},(2)}(\bb{k}_1,\bb{k}_2)
\end{equation}
depends only on $b_{2,1}^{\text{g}}$. This last degeneracies would be broken by considering higher multipoles. If we considered the tensor power spectrum too (at one loop), there would be four additional parameters, namely $\alpha_\text{ss}^{11},\alpha_\text{gg}^{11}$ and two non-local counterterms $b_R^{\{\text{s,g}\}}$ proportional to $k^2P_L(k)$\footnote{Third-order bias parameters are typically very degenerate with non-local counterterms $b_R^{\{\text{s,g}\}}$, see e.g. \cite{ivanov_pow},\cite{bakx_eft}.}. If we only consider bispectra with at most one tensor field, the stochastic parameters $\alpha_\text{gg}^{1\delta}$ and $\alpha_{\text{sgg}}^{111}$ drop out and the bispectra depend on 10 free parameters (seven from the bias expansion and three stochastic amplitudes, as mentioned at the end of Section \ref{sec:theory}). 
\section{Angular Bases for the Bispectrum}\label{sec:angbasis}
Instead of integrating the three-dimensional bispectra against spherical harmonics, we can also integrate against Legendre polynomials of \textit{different} cosines, i.e. $\mathcal{L}_\ell(\mu_2)$ \cite{damico_bossbisp}. These can be written in terms of the spherical harmonics basis by noting that 
\begin{equation}
    \mu_2 = \mu_1 \mu_{12} + Y_1^1(\mu_1,\xi) - Y_1^{-1}(\mu_1,\xi) = \mu_{12} Y_{10}(\mu_1,\xi) - \sqrt{2}Y_{11}(\mu_1,\xi).
\end{equation}
Thus, any Legendre polynomial in $\mu_2$ is ultimately a linear combination of real spherical harmonics \textit{of cosine type}, since spherical harmonics of cosine type remain of cosine type when taking products. Conversely, every real spherical harmonic of cosine type can be written as a linear combination of Legendre polynomials of $\mu_1$ and $\mu_2$, for which we provide an argument below. As such, bispectra that are even under parity transformations can be completely described in terms of either basis. The Legendre basis is certainly more convenient, as separable estimators in terms of Legendre polynomials are easy to implement. The proof of the equivalence is a straightforward induction argument over $\ell$ for every $m$. Recall the definition of the associated Legendre polynomial 
\begin{equation}\label{eq:legdef}
    \mathcal{L}_\ell^m(\mu_1) = (-1)^m (1-\mu_1^2)^{m/2} \bigg(\frac{\md}{\md\mu_1}\bigg)^m\mathcal{L}_\ell(\mu_1)
\end{equation}
for $m\geq 0$, and the relation 
\begin{equation}
    \mathcal{L}_\ell^{-m}(\mu_1) = (-1)^m\frac{(\ell-m)!}{(\ell+m)!}\mathcal{L}_\ell^m(\mu_1)
\end{equation}
extends the definition to $m<0$. Note that when $m$ is fixed, $\mathcal{L}_\ell^m$ is only nonzero for $\ell \geq |m|$. To see how a spherical harmonic of cosine type can be written as a linear combination of polynomials of $\mu_1$ and $\mu_2$, consider $m>0$ to be fixed. We can use the explicit expression
\begin{equation}\label{eq:sphdef}
    Y_{\ell m}(\mu_1,\xi) = (-1)^m \sqrt{\frac{2\ell+1}{2\pi}\frac{(\ell-m)!}{(\ell+m)!}}\mathcal{L}_\ell^m(\mu_1)\cos{(m\,\xi)} = (-1)^m c_{\ell m} \mathcal{L}_\ell^m(\mu_1)\cos{(m\,\xi)}
\end{equation}
as well as a recursion relation for associated Legendre polynomials of the form 
\begin{equation}
    (\ell-m+1)\mathcal{L}_{\ell+1}^m(\mu_1) = (2\ell+1)\mu_1 \mathcal{L}_\ell^m(\mu_1)-(\ell+m)\mathcal{L}_{\ell-1}^m(\mu_1)
\end{equation}
(which is also valid for $\ell=m$ and $\ell=m-1$) to see that for some nonzero constants $a_{\ell m},a'_{\ell m},a''_{\ell m}$,
\begin{equation}\label{eq:recur}
    a_{\ell m}Y_{\ell+1\,m}(\mu_1,\xi) = a'_{\ell m}\mu_1 Y_{\ell m}(\mu_1,\xi) + a''_{\ell m}Y_{\ell-1\,m}(\mu_1,\xi).
\end{equation}
Thus, if the claim holds for all $\ell'\leq \ell$, then it also holds for $\ell'=\ell+1$. It therefore suffices to show that the claim holds for $\ell=m$. In that case, we can in fact write 
\begin{equation}
    Y_{mm}(\mu_1,\xi) = c_{mm}(2m-1)!!\bigg(\sqrt{1-\mu_1^2}\bigg)^m \cos{(m\,\xi)}. 
\end{equation}
But $\cos{(m\,\xi)}$ is just a polynomial in $\cos \xi$ of degree $m$ (more precisely, $\cos{(m\,\xi)} = T_m(\cos \xi)$ where $T_m$ is a Chebyshev polynomial of the first kind). Moreover, it is even or odd depending on the parity of $m$. Thus, if we write 
\begin{equation}
    \cos{(m\,\xi)} = w_m (\cos \xi)^m + w_{m-2}(\cos \xi)^{m-2}+ \dots,
\end{equation}
then we immediately get 
\begin{equation}
    Y_{mm}(\mu_1,\xi) = c_{mm}(2m-1)!!\bigg[w_m\bigg(\sqrt{1-\mu_1^2}\cos\xi\bigg)^m+(1-\mu_1^2)w_{m-2}\bigg(\sqrt{1-\mu_1^2}\cos\xi\bigg)^{m-2} + \dots \bigg].
\end{equation}
However, we also know that\footnote{Throughout this argument, we ignore any issues that may arise when $\mu_{12}=\pm1$ which is the case for degenerate triangles.} 
\begin{equation}
    \sqrt{1-\mu_1^2}\cos\xi = \frac{\mu_2-\mu_{12}\mu_1}{\sqrt{1-\mu_{12}^2}}
\end{equation}
so that the statement is proved for $Y_{mm}$. This argument also shows that using all of the real spherical harmonics of cosine type up to a given $\ell$ is equivalent to using all combinations of Legendre polynomials up to a total degree $\ell$: if the $Y_{\ell m}$ on the RHS of Eq. \eqref{eq:recur} are at most of total degree $\ell$ in $\mu_1$ and $\mu_2$, then the LHS is at most of degree $\ell+1$. In summary, the set of multipoles $B^{\ell m}$ from Eq. \eqref{eq:evenmult} is equivalent to the set of OLMs $B^{O,\ell_1\ell_2}$ from Eq. \eqref{eq:newbasis_even} for parity-even bispectra. There are $(\ell+1)(\ell+2)/2$ multipoles to be formed up to total degree $\ell$. For example, for $\ell_\text{max}=2$ there are six multipoles: one monopole, two dipoles with $(\ell_1,\ell_2) = (0,1), (1,0)$ and three quadrupoles with $(\ell_1,\ell_2) = (2,0), (1,1), (0,2)$ (see \cite{damico_bossbisp} for an equivalent enumeration of the three quadrupoles).  

Additionally, if we consider the parity-odd bispectra expanded in the sine type basis, we can write for $m<0$ (and for slightly different $c_{\ell m}$ than the ones above)
\begin{equation}
    Y_{\ell m}(\mu_1,\xi) = (-1)^m \sqrt{\frac{2\ell+1}{2\pi}\frac{(\ell-|m|)!}{(\ell+|m|)!}}\mathcal{L}_\ell^{|m|}(\mu_1)\sin{(|m|\,\xi)} = (-1)^m c_{\ell m} \mathcal{L}_\ell^{|m|}(\mu_1)\sin{(|m|\,\xi)}.
\end{equation}

The first parity-odd multipole $B^{\ell m}$ appearing is $B^{1\,-1}$ since
\begin{equation}
    \hat{\bb{n}} \cdot (\hat{\bb{q}}_1 \times \hat{\bb{q}}_2) = \sqrt{(1-\mu_1^2)(1-\mu_{12}^2)}\sin \xi  = -\sqrt{(1-\mu_{12}^2)}\sqrt{\frac{4\pi}{3}}Y_{1\,-1}(\mu_1,\xi).
\end{equation}
We can show that all the other real spherical harmonic multipoles $B^{\ell m}$ with $m<0$ can be constructed by taking linear combinations of Legendre polynomials of $\mu_1$ and $\mu_2$  multiplied by $Y_{1\,-1}$. The proof is analogous, so we start by taking $m<0$ to be fixed. After using the same recursion relation for the associated Legendre polynomials again, we conclude that we need to prove the claim for $\ell=|m|$. We now express $\sin{(|m|\,\xi)}$ as $\sin \xi$ multiplied by a polynomial in $\cos \xi$ of lesser degree (note $|m|\geq 1$; and this polynomial is the Chebyshev polynomial of the second kind $U_{|m|-1}(x)$):
\begin{equation}
    \sin{(|m|\,\xi)} = \sin \xi\bigg( w_{|m|-1} (\cos \xi)^{|m|-1} + w_{|m|-3}(\cos \xi)^{|m|-3}+ \dots\bigg)
\end{equation}
Thus, 
\begin{equation}
\begin{aligned}
    Y_{|m|m}(\mu_1,\xi) &= c_{|m||m|}(2|m|-1)!!\bigg(\sqrt{1-\mu_1^2}\bigg)^{|m|} \sin{(|m|\,\xi)} \\
    &= c_{|m||m|}(2|m|-1)!!\bigg(\sqrt{1-\mu_1^2} \sin \xi\bigg) \\
    &\times \bigg[w_{|m|-1}\bigg(\sqrt{1-\mu_1^2}\cos\xi\bigg)^{|m|-1}+(1-\mu_1^2)w_{|m|-3}\bigg(\sqrt{1-\mu_1^2}\cos\xi\bigg)^{|m|-3} + \dots \bigg] \\
\end{aligned}
\end{equation}
where we recognize $Y_{1\,-1}$ on the second line, and the expression involving Legendre polynomials on the third. Thus, considering all spherical harmonics of sine type up to a given $\ell$ is equivalent to considering all products of $Y_{1\,-1}$ with Legendre polynomials of total degree $\ell-1$, and there are $\ell(\ell+1)/2$ such combinations. Thus, the parity odd multipoles $B^{\ell m}$ from Eq. \eqref{eq:oddmult} are completely equivalent to the parity-odd OLMs $B^{O,\ell_1\ell_2}$ from Eq. \eqref{eq:newbasis_odd}\footnote{Coincidentally, some of these low-order multipoles may vanish for the redshift-space galaxy bispectrum, but merely as a consequence of the Newtonian approximation. Specifically, multipoles with odd values of $\ell$ (i.e. an odd total degree of Legendre polynomials) do not occur, such as the dipole terms $(\ell_1,\ell_2) = (1,0)$ or $(\ell_1,\ell_2) = (0,1)$. However, when one includes relativistic or wide-angle effects, these terms do appear \cite{deweerd_bisp}.}. 

While these arguments show that the above expansions are complete, it may still be that some multipoles are mathematically identical in case two of the sides are equal or some of the tracers are. We elaborate on this in Section \ref{sec:forecast}. 
\section{Symbolic Angular Integration}\label{sec:angint}
In our implementation, we found it practical where possible to rewrite all symbolic integrations necessary for the computation of the multipoles themselves and the covariance matrices in terms of the angular variables $\mu_1$ and $\xi$. Then, one can proceed by using 
\begin{equation}
    I_{n_1, n_2} = \int_{-1}^1\md\mu_1\big(\sqrt{1-\mu_1^2}\big)^{n_1-1}\mu_1^{n_2} = \int_{0}^\pi \md \theta_1 \sin^{n_1}\theta_1 \cos^{n_2}\theta_1 = (1+(-1)^{n_2})\frac{\Gamma(\frac{1+n_1}{2})\Gamma(\frac{1+n_2}{2})}{2\Gamma(\frac{2+n_1+n_2}{2})}.
\end{equation}
so that
\begin{equation}
    \frac{1}{4\pi}\int\md\mu_1\md\xi\, \bigg(\big(\sqrt{1-\mu_1^2}\big)^{n_1}\mu_1^{n_2}\sin^{m_1}\xi \cos^{m_2}\xi\bigg) = \frac{1}{4\pi}I_{n_1+1,n_2}I_{m_1,m_2}(1 +(-1)^{m_1}).
\end{equation}
\section{Spectra with Parity-odd Fields}\label{sec:parity}
Some of our results regarding non-vanishing three-point statistics with an odd number of B-modes may appear to be at odds with current literature. For example, \cite{schneider_parity,pyne_illustris} claim that such spectra should vanish based on symmetry arguments. However, as has already been noted in e.g. \cite{schmitz_bisp} and in our main text, in contrast to two-point statistics, performing a reflection in a plane containing the line of sight does not in general map a bispectrum triangle onto itself. Therefore, the relation that is inferred from parity symmetry reads (taking $B_{DDB}$ as an example)
\begin{equation}
    B_{DDB}(\bb{k}_1',\bb{k}_2',\bb{k}_3') = - B_{DDB}(\bb{k}_1,\bb{k}_2,\bb{k}_3)
\end{equation}
rather than $B_{DDB}(\bb{k}_1,\bb{k}_2,\bb{k}_3)= - B_{DDB}(\bb{k}_1,\bb{k}_2,\bb{k}_3)$ where the latter would imply that the bispectrum vanishes. By contrast, the power spectrum $P_{D B}(k,\mu)$ actually only depends on the magnitude of the wavevector and its angle with the line of sight $\mu$, both of which do not change under a reflection $\bb{P}$. Thus, we do have the implication 
\begin{equation}
    P_{D B}(\bb{k}) = P_{D B}(\bb{k}') = -P_{D B}(\bb{k}) \implies P_{DB}(\bb{k}) =0. 
\end{equation}
There is an additional caveat: even though odd three-dimensional bispectra do not vanish (but rather are odd under reflections), an \textit{angular average} of such an odd function may vanish. For example, we saw that 
\begin{equation}
    B_{DDB}^{\ell m}(k_1,k_2,k_3) =
    \frac{2\ell+1}{4\pi}\int\md\mu_1\md\xi\, B_{DDB}(k_1,k_2,k_3,\mu,\xi) Y_{\ell m}(\mu,\xi) = 0
\end{equation}
for any positive $m$, because of the vanishing integral over $\xi$. It is beyond the scope of this work to write down explicit expressions for general line-of-sight-integrated bispectra and other three point statistics such as the aperture mass \cite{linke_lowz,shear_3pt1,shear_3pt2,shear_3pt3}. However, we remark that null measurements of `odd' statistics such as $\langle M_\text{ap}^2(\theta) M_{\times}(\theta) \rangle$ need to be interpreted with caution when attempting to translate them to conclusions about the three-point statistic inside the integrand. An integral of an odd statistic may also vanish if the statistic is evaluated in a `symmetric' configuration, such as a collinear or equilateral triangle (see the discussion in Section \ref{sec:results}). 

\section{Gaussian Covariance}\label{sec:cov}
Here we compute the Gaussian contribution to the bispectrum covariance. The binned estimator for the standard bispectrum multipoles is given by 
\begin{equation}\label{eq:est}
    \hat{B}^{T,\ell_1\ell_2}_{XYZ}(k_1,k_2,k_3) = \frac{N_{\ell_1\ell_2}}{N_\Delta V}\sum_{\mathbf{q}_i \in \text{bin } i}\delta^K(\mathbf{q}_{123}) \hat{X}(\mathbf{q}_1) \hat{Y}(\mathbf{q}_2)\hat{Z} (\mathbf{q}_3)\mathcal{P}_{XYZ}^{T,\ell_1\ell_2}(\{\hat{\bb{q}}_i\}).
\end{equation}
Hence, the sum runs over all the wavenumber triplets that fall within the specified bin and sum to zero. The normalization factor $N_\Delta$ equals the number of fundamental triangles present in the $(k_1,k_2,k_3)$ bin. We do not order the bins. The quantity $\hat{X}$ refers to the \textit{measured} field, i.e. it is evaluated on a discrete grid by means of FFTs.The Kronecker delta symbol $\delta^K$ equals unity when its argument is zero and zero otherwise. To compute the sum above in practice, we use the FFT-based estimator from \cite{scoccimarro_fast}. The measured fields can be related to the theory spectra via the relations\footnote{In the continuum limit, we have $V\delta^K(\bb{q}) \to (2\pi)^3\delta^D(\bb{q})$ so that this definition is consistent with our theoretical definition of the correlators (cf. Eq. \eqref{eq:bispdef}).} 
\begin{equation}\label{eq:dspec}
\begin{aligned}
    \langle \hat{X}(\bb{q}_1)\hat{Y}(\bb{q}_2)\rangle &= V \delta^K(\bb{q}_{12}) P_{XY}(\bb{q}_1); \\
    \langle \hat{X}(\bb{q}_1)\hat{Y}(\bb{q}_2)\hat{Z}(\bb{q}_3)\rangle &= V \delta^K(\bb{q}_{123}) B_{XYZ}(\bb{q}_1,\bb{q}_2,\bb{q}_3). \\
\end{aligned}
\end{equation}
As usual, one can `realize' the ensemble average $\langle \cdot \rangle$ over initial conditions by generating a sufficient number of independent simulations. By using the second line of Eq. \eqref{eq:dspec}, we see that the ensemble average of the estimator equals 
\begin{equation}\label{eq:multipint}
\begin{aligned}
    \langle \hat{B}^{T,\ell_1\ell_2}_{XYZ}(k_1,k_2,k_3)\rangle &= \frac{N_{\ell_1\ell_2}}{N_\Delta}\sum_{\mathbf{q}_i \in \text{bin } i}\delta^K(\mathbf{q}_{123}) B_{XYZ}(\bb{q}_1,\bb{q}_2,\bb{q}_3) \mathcal{P}_{XYZ}^{T,\ell_1\ell_2}(\{\hat{\bb{q}}_i\}) \\
    &\approx \frac{N_{\ell_1\ell_2}}{V_\Delta}\int_{k_1}\md^3\bb{q}_1\int_{k_2}\md^3\bb{q}_2\int_{k_3}\md^3\bb{q}_3 \,\delta^D(\mathbf{q}_{123})B_{XYZ}(\bb{q}_1,\bb{q}_2,\bb{q}_3)\mathcal{P}_{XYZ}^{T,\ell_1\ell_2}(\{\hat{\bb{q}}_i\}) \\
    &\approx \frac{N_{\ell_1\ell_2}}{4\pi}\int \md\mu_1\md\xi\, B_{XYZ}(k_1,k_2,k_3,\mu_1,\xi)\mathcal{P}_{XYZ}^{T,\ell_1\ell_2}(\mu_1,\xi),
\end{aligned}
\end{equation}
where in the second line we used the continuous approximation to make the replacement 
\begin{equation}
    \frac{1}{N_\Delta}\sum_{\mathbf{q}_i \in \text{bin } i}\delta^K(\mathbf{q}_{123}) \approx \frac{1}{V_\Delta}\int_{k_1}\md^3\bb{q}_1\int_{k_2}\md^3\bb{q}_2\int_{k_3}\md^3\bb{q}_3\, \delta^D(\mathbf{q}_{123});
\end{equation}
with $V_\Delta$ the volume of the integration region \footnote{Note the factor of $(2\pi)^6$ difference with e.g. \cite{damico_bossbisp}.}. In the third line we employed the thin-shell approximation so that the radial integrals over the bins become trivial and the angular average is a spherical integral, viz. for an arbitrary rotationally invariant integrand $U(\bb{q}_1,\bb{q}_2,\bb{q}_3)$ (that is, an integrand only depends on the orientation of the triangle with respect to the line of sight) we assert that \cite{mehrem1}
\begin{equation}
\begin{aligned}
    \frac{1}{V_\Delta}\int_{k_1}\md^3\bb{q}_1&\int_{k_2}\md^3\bb{q}_2\int_{k_3}\md^3\bb{q}_3 \, \delta^D(\mathbf{q}_{123})\, U(\bb{q}_1,\bb{q}_2,\bb{q}_3) \approx \frac{1}{4\pi}\int\md\mu_1\md\xi \,U(k_1,k_2,k_3,\mu_1,\xi).
\end{aligned}
\end{equation} 
We can also use (cf. Eq. \eqref{eq:cross2})
\begin{equation}\label{eq:cross}
    \hat{\bb{n}} \cdot (\hat{\bb{q}}_1 \times \hat{\bb{q}}_2) = \sqrt{(1-\mu_1^2)(1-\mu_{12}^2)}\sin \xi 
\end{equation}
to express an odd projector in terms of $\mu_1$ and $\xi$ in the bottom line of Eq. \eqref{eq:multipint}.

In a similar vein, the covariance can also be calculated: we can correlate two arbitrary instances of Eq. \eqref{eq:est} and subtract the product of their means to get
\begin{equation}\label{eq:covcalc}
\begin{aligned}
    \langle \hat{B}&^{T,\ell_1 \ell_2}_{XYZ}(k_1,k_2,k_3) \hat{B}^{T,\ell_1' \ell_2'}_{X'Y'Z'} (k_1',k_2',k_3')\rangle - \langle \hat{B}^{T,\ell_1 \ell_2}_{XYZ}(k_1,k_2,k_3) \rangle \langle \hat{B}^{T,\ell_1' \ell_2'}_{X'Y'Z'} (k_1',k_2',k_3')\rangle \\
    &= \frac{N_{\ell_1\ell_2}N_{\ell_1'\ell_2'}}{N_\Delta N_{\Delta'}V^2}\sum_{\mathbf{q}_i \in \text{bin } i}\sum_{\mathbf{q}_j' \in \text{bin } j}\delta^K(\mathbf{q}_{123})\delta^K(\mathbf{q}'_{123}) \times\bigg( \langle \hat{X}(\mathbf{q}_1) \hat{Y}(\mathbf{q}_2)\hat{Z}(\mathbf{q}_3)\hat{X'}(\mathbf{q}'_1)\hat{Y'}(\mathbf{q}'_2)\hat{Z'} (\mathbf{q}'_3) \rangle \\
    &- \langle \hat{X}(\mathbf{q}_1) \hat{Y}(\mathbf{q}_2)\hat{Z}(\mathbf{q}_3)\rangle \langle \hat{X'}(\mathbf{q}'_1)\hat{Y'}(\mathbf{q}'_2)\hat{Z'} (\mathbf{q}'_3) \rangle  \bigg)\bigg(\mathcal{P}_{XYZ}^{T,\ell_1\ell_2}(\{\hat{\bb{q}}_i\})\mathcal{P}_{X'Y'Z'}^{T,\ell_1'\ell_2'}(\{\hat{\bb{q}}'_i\})\bigg).
\end{aligned}
\end{equation}
To simplify this, we will need to take a number of steps. The first term on the last line is a six-point function, so that we can expand the term in parentheses as \footnote{Note that Wick's theorem in principle also generates `self-terms' such as $P_{XY}(\bb{q}_1)P_{ZX'}(\bb{q}_3)P_{Y'Z'}(\bb{q}_2')\delta^K(\bb{q}_1+\bb{q}_2)\delta^K(\bb{q}_3+\bb{q}'_1)\delta^K(\bb{q}'_2+\bb{q}'_3)$, but these enforce e.g. $\bb{q}_1+\bb{q}_2 = 0$ and thus also $\bb{q}_3 = 0$, since the estimator involves only the \textit{diagonal}, i.e. `on-shell' bispectrum. Such configurations contribute negligibly to the sum and will hence be ignored. } 
\begin{equation}\label{eq:wick}
\begin{aligned}
    V^{-3}& \bigg(\langle  \hat{X}(\mathbf{q}_1) \hat{Y}(\mathbf{q}_2)\hat{Z}(\mathbf{q}_3)\hat{X'}(\mathbf{q}'_1)\hat{Y'}(\mathbf{q}'_2)\hat{Z'} (\mathbf{q}'_3) \rangle 
    - \langle \hat{X}(\mathbf{q}_1) \hat{Y}(\mathbf{q}_2)\hat{Z}(\mathbf{q}_3)\rangle \langle \hat{X'}(\mathbf{q}'_1)\hat{Y'}(\mathbf{q}'_2)\hat{Z'} (\mathbf{q}'_3) \rangle\bigg) \\
    &= P_{XX'}(\bb{q}_1)P_{YY'}(\bb{q}_2)P_{ZZ'}(\bb{q}_3)\delta^K(\bb{q}_{11'})\delta^K(\bb{q}_{22'})\delta^K(\bb{q}_{33'}) 
    +P_{XX'}(\bb{q}_1)P_{YZ'}(\bb{q}_2)P_{ZY'}(\bb{q}_3)\delta^K(\bb{q}_{11'})\delta^K(\bb{q}_{23'})\delta^K(\bb{q}_{32'}) \\
    &+ P_{XY'}(\bb{q}_1)P_{YX'}(\bb{q}_2)P_{ZZ'}(\bb{q}_3)\delta^K(\bb{q}_{12'})\delta^K(\bb{q}_{21'})\delta^K(\bb{q}_{33'}) 
    + P_{XY'}(\bb{q}_1)P_{YZ'}(\bb{q}_2)P_{ZX'}(\bb{q}_3)\delta^K(\bb{q}_{12'})\delta^K(\bb{q}_{23'})\delta^K(\bb{q}_{31'}) \\
    &+ P_{XZ'}(\bb{q}_1)P_{YY'}(\bb{q}_2)P_{ZX'}(\bb{q}_3)\delta^K(\bb{q}_{13'})\delta^K(\bb{q}_{22'})\delta^K(\bb{q}_{31'}) 
    + P_{XZ'}(\bb{q}_1)P_{YX'}(\bb{q}_2)P_{ZY'}(\bb{q}_3)\delta^K(\bb{q}_{13}')\delta^K(\bb{q}_{21'})\delta^K(\bb{q}_{32'})  \\
    &= \sum_{\sigma \in S_3}P_{X\sigma(X')}(\bb{q}_1)P_{Y\sigma(Y')}(\bb{q}_2)P_{Z\sigma(Z')}(\bb{q}_3)\delta^K(\bb{q}_{1\sigma(1')})\delta^K(\bb{q}_{2\sigma(2')})\delta^K(\bb{q}_{3\sigma(3')})
\end{aligned}
\end{equation}
where we neglect higher-order corrections such as the bispectrum-bispectrum ($BB$), power spectrum-trispectrum ($PT$) and connected pentaspectrum ($T_6$) terms \cite{sefusatti_bisp1, sefusatti_bisp2} in the covariance, as well as the product of the estimator averages. These are at least eighth order in perturbations and therefore considered subdominant. In the bottom line, we have written the expression as a sum over six permutations $\sigma \in S_3$ which is understood to act on the ordered sets $\{X',Y',Z'\}$ and $\{1,2,3\})$ simultaneously. We can conclude that the covariance vanishes if the triangles are not similar, i.e. if $(k_1',k_2',k_3')$ is not a permutation of $(k_1,k_2,k_3)$. After inserting the bottom line of Eq. \eqref{eq:wick} into Eq. \eqref{eq:covcalc} we can carry out the sums over the primed variables. For any permutation $\sigma$, the occurrence of $\delta^K(\bb{q}_{a\sigma(a')})$ with $a=1,2$ or $3$ implies that we have to replace every occurrence of the primed vector $\bb{q}'_{\sigma(a)}$ by $-\bb{q}_{a}$, or equivalently we should replace every $\bb{q}'_{a}$ by $-\bb{q}_{\bar{\sigma}(a)}$ where $\bar{\sigma}$ is the inverse permutation of $\sigma$. This is especially relevant for the projector terms in Eq. \eqref{eq:covcalc}, which contain unit vectors as their arguments. As such, these arguments need to be interchanged consistently. We obtain
\begin{equation}\label{eq:covresult}
\begin{aligned}
    \langle& \hat{B}^{T,\ell_1 \ell_2}_{XYZ}(k_1,k_2,k_3) \hat{B}^{T,\ell_1' \ell_2'}_{X'Y'Z'} (k_1',k_2',k_3')\rangle - \langle \hat{B}^{T,\ell_1 \ell_2}_{XYZ}(k_1,k_2,k_3) \rangle \langle \hat{B}^{T,\ell_1' \ell_2'}_{X'Y'Z'} (k_1',k_2',k_3')\rangle \\
    &= \frac{N_{\ell_1\ell_2}N_{\ell_1'\ell_2'}V}{N_\Delta^2}\sum_{\mathbf{q}_i \in \text{bin } i}\delta^K(\mathbf{q}_{123})\times \bigg(\sum_{\sigma \in S_3}P_{X\sigma(X')}(\bb{q}_1)P_{Y\sigma(Y')}(\bb{q}_2)P_{Z\sigma(Z')}(\bb{q}_3) \\
    &\times \delta_{1\sigma(1)'}\delta_{2\sigma(2)'}\delta_{3\sigma(3)'} \mathcal{P}_{XYZ}^{T,\ell_1\ell_2}(\{\hat{\bb{q}}_i\})\mathcal{P}_{X'Y'Z'}^{T,\ell_1'\ell_2'}(\bar{\sigma}(\{-\hat{\bb{q}}_i\}))\bigg) \\
    &\approx \frac{N_{\ell_1\ell_2}N_{\ell_1'\ell_2'}V}{N_\Delta V_\Delta } \int_{k_1}\md^3\bb{q}_1\int_{k_2}\md^3\bb{q}_2\int_{k_3}\md^3\bb{q}_3\,\delta^D(\mathbf{q}_{123}) \times \bigg(\sum_{\sigma \in S_3}P_{X\sigma(X')}(\bb{q}_1)P_{Y\sigma(Y')}(\bb{q}_2)P_{Z\sigma(Z')}(\bb{q}_3) \\
    &\times \delta_{1\sigma(1)'}\delta_{2\sigma(2)'}\delta_{3\sigma(3)'} \mathcal{P}_{XYZ}^{T,\ell_1\ell_2}(\{\hat{\bb{q}}_i\})\mathcal{P}_{X'Y'Z'}^{T,\ell_1'\ell_2'}(\bar{\sigma}(\{\hat{\bb{q}}_i\}))\bigg)  \\
    &\approx \frac{N_{\ell_1\ell_2}N_{\ell_1'\ell_2'}V}{4\pi N_\Delta} \int \md\mu_1\md\xi  \,\bigg(\sum_{\sigma \in S_3}P_{X\sigma(X')}(k_1,\mu_1)P_{Y\sigma(Y')}(k_2,\mu_2)P_{Z\sigma(Z')}(k_3,\mu_3) \\
    &\times \delta_{1\sigma(1)'}\delta_{2\sigma(2)'}\delta_{3\sigma(3)'} \mathcal{P}_{XYZ}^{T,\ell_1\ell_2}(\{\hat{\bb{q}}_i\})\mathcal{P}_{X'Y'Z'}^{T,\ell_1'\ell_2'}(\bar{\sigma}(\{\hat{\bb{q}}_i\}))\bigg)
\end{aligned}    
\end{equation}
where we used the continuous approximation once again. The unit vectors are given by 
\begin{equation}
    \begin{aligned}
        \hat{\bb{q}}_1 &= (\sqrt{1-\mu_1^2},0,\mu_1); \\
        \hat{\bb{q}}_2 &= (\mu_1\sqrt{1-\mu_{12}^2}\cos \xi + \mu_{12}\sqrt{1-\mu_1^2},\sqrt{1-\mu_{12}^2}\sin \xi,\mu_1\mu_{12}-\sqrt{1-\mu_{12}^2}\sqrt{1-\mu_{1}^2}\cos\xi); \\
        \hat{\bb{q}}_3 &= \frac{q_1\hat{\bb{q}}_1}{q_3}-\frac{q_2\hat{\bb{q}}_2}{q_3}.
    \end{aligned}
\end{equation}
as in Eq. \eqref{eq:unitvecs} (but now $\mu_{12}$ is the angle between $\bb{q}_1$ and $\bb{q}_2$). The symbol $\delta_{aa'}$ equals unity if $k_a = k'_{a}$ and zero otherwise. Note that the volume of the region of integration $V_\Delta$ satisfies $N_\Delta \approx k_f^{-6}V_\Delta $ since the density of modes in Fourier space is $k_f^{-3}$. For the results in Section \ref{sec:results}, we calculate this volume via explicit integration over the radial bins; not doing so would result in large errors for collinear configurations. 
As expected, the covariance is inversely proportional to the volume of the simulation (provided, of course, that the number density and hence the shape noise stays fixed). The bottom line of Eq. \eqref{eq:covresult} is symmetric under exchanging all primed and unprimed quantities, as it should be. 

As a final remark, because Eq. \eqref{eq:covresult} is completely general in terms of tracers and wavenumber configurations, it is actually quite convenient to implement. We can use it to compute all auto- and cross covariances of tensorial bispectra in the Gaussian approximation. In the cases where Eq. \eqref{eq:covresult} does not evaluate to zero, the sum over permutations has $6,2$ or $1$ nonzero terms depending on whether the triangle is equilateral, isosceles or general, respectively. There is no cross-covariance between parity-even and parity-odd bispectra in the Gaussian approximation, since power spectra with an odd number of B-modes always vanish.
\bibliographystyle{JHEP}
\newpage
\bibliography{mybibliography}
\end{document}